\documentclass[twocolumn,showpacs,preprintnumbers,amsmath,amssymb]{revtex4}
\usepackage{dcolumn}
\usepackage{bm}
\usepackage{verbatim}
\usepackage{graphicx}

\newcommand{\kb}{k_{\scriptsize B}}
\newcommand{\ud}{\mathrm{d}}
\newcommand{\lco}{\ell}
\newcommand{\lcoq}{\ell^2}
\newcommand{\fulle}{{C_{60}}}
\newcommand{\tf}{T}
\newcommand{\LT}{L_T}
\newcommand{\tT}{t_T}

\begin{document}

\title{Analysis of the loss of coherence in interferometry with
  macromolecules }

\author{A. Viale}\email{viale@ge.infn.it} \author{M.
  Vicari}\email{vicari@fisica.unige.it} \author{N.
  Zangh\`\i}\email{zanghi@ge.infn.it}
\homepage{http://www.ge.infn.it/~zanghi} \affiliation{Dipartimento di
  Fisica, Istituto Nazionale di Fisica Nucleare, Sezione di Genova,
  Via Dodecaneso 33, 16146 Genova, Italy}

\date{\today}

\begin{abstract}
  We provide a self-contained quantum description of the interference
  produced by macromolecules diffracted by a grating, with particular
  reference to fullerene interferometry
  experiments~\cite{arnzei99,JMO2000}. We analyze the processes
  inducing loss of coherence consisting in beam preparation
  (collimation setup and thermal spread of the wavelengths of the
  macromolecules) and in environmental disturbances. The results show
  a good agreement with experimental data and highlight the analogy
  with optics. Our analysis gives some hints for planning future
  experiments.
\end{abstract}

\pacs{03.65.Yz, 03.75.-b, 03.65.Ta, 39.20.+q}

\maketitle

\section{Introduction}

The aim of this paper is to provide a theoretical derivation of the
measured beam intensity profile for interferometry with macromolecular
beams, under the influence of the processes inducing loss of
coherence, consisting both in beam preparation (collimation setup and
thermal spread of the wavelengths of the macromolecules) and in
environmental disturbances.

We have been motivated by the impressive experiments with fullerene
made by Zeilinger's group~\cite{arnzei99,JMO2000}. In these
experiments, thermally produced beams of heavy macromolecules are
collimated, diffracted by a grating and then detected on a distant
screen. The diffraction pattern so produced shows the typical
interference profile of wave phenomena in the presence of incoherent
contributions, and the reduced fringe visibility observed reminds
very much of Kirchhoff diffraction with thermal light produced by an
extended source. By taking into account the effects of the
interaction of the macromolecules with the environment and with the
photons they emit by internal cooling, we shall provide a
self-contained {\em
  quantum} description of these experiments, that does not rely on
methods of classical optics.

Our analysis is based on two main ingredients.  The first one is a
matter of principle: it is the formula for the statistics of particle
arrival position and time on a distant surface given by
Eq.~\eqref{basicequation} below.  According to this formula, the
intensity pattern revealed on a distant screen can be expressed in
terms of the large-distance asymptotic behavior of the
time-integrated quantum current (see Eq.~\eqref{curint} below).  This
formula was conjectured long time ago within the framework of
scattering theory~\cite{Combes}, and only recently has been
derived~\cite{nino2,
  det2, det3}, extended to the mesoscopic regime, and physically
motivated within the framework of Bohmian mechanics~\cite{nino}.

The second ingredient is a matter of analysis to simplify the dynamics
of a test particle moving in a quantum medium: it is the model of Joos
and Zeh~\cite{joze85} for the phenomenological description of
processes inducing loss of coherence in quantum systems. In this
model, the reduced density matrix of the system evolves autonomously
according to a ``Boltzmann-type'' master equation. The effect of the
environment is summarized by ``a collision term'', added to the free
dynamics of the system, which takes into account the
\emph{decoherence}, i.e., damping of the off-diagonal terms of the
density matrix in position representation.

Starting from these premises, and by means of some approximations that
are reasonable in the common experimental conditions for
interferometry with heavy particles, we shall derive an easy relation
useful to describe diffraction patterns.

Finally, we shall provide a theoretical fit for the experimental data
reported in~\cite{JMO2000}, and we shall discuss some predictions
about the dependence of the interference patterns on physical
parameters such as the mass of the molecules, the pressure at which
the experiment is performed and the distance of the detection screen.
These predictions can be of some relevance for planning new
experiments.

\section{Measured Intensity}
\label{section32}

In the experiments with fullerenes a thermally produced beam of heavy
macromolecules is collimated, diffracted by a grating and then
detected on a distant screen (see Fig.~\ref{fig1}). The grating is
composed by parallel slits, and it is periodic of period $D$. The
grating and the detection screen lie in parallel planes, which are
orthogonal to the longitudinal direction of beam propagation ($y$
direction), the so-called {\em optical axis}. The screen is placed at
the distance $L$ from the grating. During the flight from the grating
to the screen the fullerenes interact with air at low pressure, as
well as with thermal photons, and get entangled with the photons
emitted by relaxation of the internal excited states.  What is the
theoretical prediction for the particle intensity measured at the
screen? In order to answer to this question we shall proceed in two
steps: firstly, in Section~\ref{sec2}, we shall consider the case in
which the dynamics of center of mass of the diffracted particle is
governed by free Schr\"odinger evolution. Though this approximation
is unrealistic for fullerenes, it is a good approximation for lighter
particles as electrons or neutrons.  Then, starting with Section
\ref{iefi2}, we shall refine the description and take into account
the influence of the environment on the motion of the particle.

\subsection{Free evolution}\label{sec2}

In a typical detection experiment the count statistics is obtained by
summing a large number of events in which the particle crosses the
detection screen at a random time. What is the appropriate quantum
prediction for such a statistics?  This question is not quite as
innocent as it sounds; it concerns in fact one of the most debated
problems in quantum theory, the problem of time measurement,
specifically the problem of arrival time, and position at such time.
It is well known that there is no self-adjoint time observable of any
sort, and since the arrival position is the position of the particle
at a random time, it cannot be expressed as a Heisenberg position
operator in any obvious way.  Bohmian mechanics does provide, however,
a remarkably simple answer (for an updated review of Bohmian mechanics
see~\cite{BM1} and references therein): Let $S$ be a surface in
physical space, $\Psi({\bf r},t)$ be the wave function of a particle,
and
 \begin{equation}\nonumber
\mathbf{J}({\bf r},t) =\frac{\hbar}{M}\,{\rm Im}\left[\Psi({\bf r},t)^*
 \:\nabla_{\bf r}\Psi({\bf r},t)\right]
 \label{basicflux}
\end{equation}
be the associated quantum current satisfying the continuity equation
 \begin{equation}\nonumber
\frac{\partial |\Psi({\bf r},t)|^2}{\partial t} + \nabla_{\bf r}
\cdot {\bf J({\bf r},t)} = 0.
 \label{continuityequation}
\end{equation}

Then the joint probability $\mathsf{Prob}({\bf R}_{T}\in dS, T\in dt)$
that the particle crosses the surface element $d{\bf S}$ of the
surface $S$ at the point ${\bf r}$ in the time between $t$ and $t+dt$
is given by
 \begin{equation}
 \label{basicequation}
\mathsf{Prob}({\bf R}_{T}\in dS, T\in dt)=
 {\bf J}({\bf r},t) \cdot d{\bf S}\, dt
\end{equation}
\emph{provided} the current positivity condition ${\bf J}({\bf r},t)
\cdot d{\bf S}> 0$ is satisfied (a condition on both the wave function
$\Psi$ and on the surface $S$). See~\cite{nino} for a general
derivation of this (for applications to mesoscopic physics
see~\cite{leavens1,leavens2,leavens3}, in this regard, see
also~\cite{grubl}).

\begin{figure}[h!]
  \includegraphics [scale=0.45]{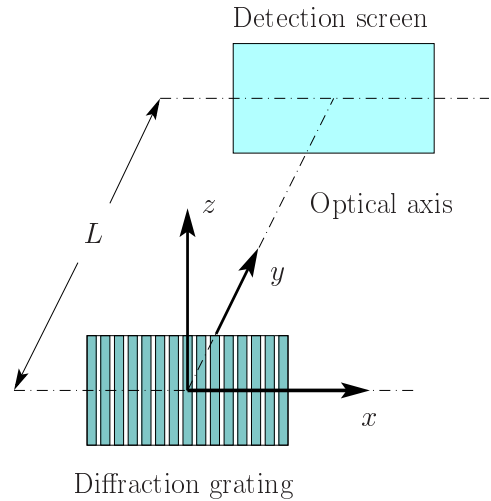}
\caption{Geometric configuration of the diffraction grating
  and the detection screen.}\label{fig1}
\end{figure}

We would like now to apply such a probabilistic prediction to the
typical situation of a diffraction experiment with particles of mass
$M$ diffracted by a grating and then detected on a distant screen.
The geometry is that of Fig.~\ref{fig1} and, by (approximate)
translational invariance along the slit extension---the $z$
axis---without loss of generality, we can consider the dynamics to be
effectively two dimensional: ${\bf r} = (x,y)$, where $x$ is the
coordinate along the grating perpendicular to the slit axes and $y$
is the coordinate perpendicular to both the grating and the detection
screen, i.e., along the so-called optical axis.

Let us make the physically reasonable assumption that the initial
($t=0$) wave function produced by the grating factorizes
\begin{equation}\label{Psi}\nonumber
\Psi({\bf r},0)=\psi_0(x)\,\phi_0(y),
\end{equation}
with the size $\Delta x$ of the support of $\psi_0$ being that of the
grating. Then $\Psi$ evolves freely according to Schr\"odinger
equation
\begin{equation}\label{sch}
i\hbar\frac{ \partial\Psi}{\partial t}= -\frac{\hbar^2}{2M}
{\nabla_{\bf r}}^2 \Psi
\end{equation}
until the particle is detected on the screen in the $xz$-plane placed
at a ``large'' distance $y=L$. This last condition---see below for a
suitable specification of how large $L$ should be---ensures that the
current positivity condition is fulfilled.  Then, according to
Eq.~\eqref{basicequation}, the probability density that the particle
crosses the screen at the point $x$ is
\begin{equation}\label{curint}
I(x)= \left.\int^{+\infty}_{0}\!\!\!\!\!\!\!\!\ud t\
 J_y(x,y,t)\right\vert_{y=L},
\end{equation}
where $J_y$ is the longitudinal component of the quantum current,
i.e.,
\begin{equation}\label{curnum}
I(x)= \frac{\hbar}{M}\int^{+\infty}_{0}\!\!\!\!\!\!\!\!\ud t
\ |\psi(x,t)|^2 \ {\rm Im}\left[ \phi(y,t)^{*}\:
\frac{\partial \phi(y,t)}{\partial y}\right]_{y=L}.
\end{equation}

For a large ensemble (beam) of particles identically prepared in the
same initial state, by the law of large numbers, $I({x})$ is
proportional to the local \emph{intensity} measured at the
screen---and without loss of generality, the proportionality constant,
which is easily determinable by the total count statistics, will be
hereafter assumed to be~$1$.

Let us now make some simplifications: Assume that the momentum
component $p_y$ is sharply defined, i.e., $\Delta p_y \ll p_y$, so
that with the initial wave function is associated a well-defined de
Broglie wavelength
\begin{equation}\label{dpp}
\lambda\sim h/p_y \ll\Delta y \,.
\end{equation}
Then we may approximate Schr\"odinger's evolution of $\phi_0$ with a
classical propagation at velocity $v=p_y/M$, so that $I(x)$ gets
approximated by
\begin{equation}\label{I1}
I_1(x) =  v\int^{+\infty}_{0}\!\!\!\!\!\!\!\!\ud t\: \big\vert\psi({x},t)
\big\vert^2\ \big\vert\phi_0\big (L-vt\big)\big\vert^2.
\end{equation}

Suppose furthermore that the detector distance $L$ is much larger than
the position spread in the longitudinal direction,
\begin{equation}\label{sL}
L\gg\Delta y.
\end{equation}
Then the time-integration in Eq.~\eqref{I1} gives appreciable
contributions only for
\begin{equation}\label{tf}
t= \tf\equiv\frac{L}{v} = \frac{M\lambda}{2\pi \hbar} L\,,
\end{equation}
which is the so-called ``time of flight,'' that is the time spent by
the particle to reach the detector. Thus, Eq.~\eqref{I1} can be
further approximated as
\begin{equation}\label{I2}
I_2({x}) =  \big\vert\psi({x}, \tf)\big\vert^2,
\end{equation}
where $\psi({x}, t)$ is the solution of one-dimensional free
Schr\"odinger's equation, i.e.,
\begin{equation}\nonumber
\psi(x,t)  =\sqrt{\frac{M}{2\pi i\hbar t}}
\int dx_0\:   e^{\frac{iM}{2\hbar t} (x-x_0)^2} \psi_0(x_0)\,.
\end{equation}

Consider now $|\psi(x,t)|^2 $, i.e.,
\begin{equation}\label{roBIS}\nonumber
\frac{M}{2\pi\hbar t}\!\iint\!\! \ud x_0 \,\ud x_0'\:
e^{\frac{iM}{2\hbar
t}[x_0^2-x_0'^2+2x(x_0'-x_0)] }
\:\psi_0(x_0)\;\psi_0(x_0')^*
\end{equation}
and note that in the integrations both $x_0$ and $x_0'$ are bounded by
$\Delta x$, the support of $\psi_0$.  Thus, if
\begin{equation}\label{condo}
\frac{M (\Delta x)^2 }{\hbar t} \ll 1
\end{equation}
we have that $ e^{\frac{iM}{2\hbar t}[x_0^2-x_0'^2] } \approx 1$ and
therefore that $|\psi(x,t)|^2 $ is approximated by
\begin{equation}\label{roTRIS}\nonumber
\frac{M}{2\pi\hbar t}\!\iint\!\! \ud x_0 \,\ud x_0'\:
e^{\frac{iM x}{\hbar t} (x_0'-x_0)  }
\psi_0(x_0)\;\psi_0(x_0')^*\,.
\end{equation}
One can easily recognize that the above expression (modulo an overall
proportionality constant) is nothing but the square of
$\widehat{\psi}_0(k)$, the Fourier transform of $\psi_0(x)$, computed
for $k=Mx/(\hbar t)$.  Therefore, by replacing in Eq.~\eqref{I2} such
an approximation of $|\psi(x,t)|^2$ for $t=\tf$, and recalling
definition \eqref{tf} for $\tf$, we arrive at a further approximation
for the intensity $I(x)$, namely,
\begin{equation}\label{I3}
I_3({x})\ =\  \frac{2\pi}{\lambda L}\
  \bigg\vert\widehat{\psi}_0\bigg (\frac{2\pi {x}}{\lambda
    L}\bigg )\bigg\vert ^2.
\end{equation}
It is important to observe that the regime in which this approximation
holds is that fixed by Eq.~\eqref{condo} for $t=\tf$ i.e., the regime
characterized by the condition
\begin{equation}\label{frau}
 \frac{\Delta x}{L}\ll \frac{\lambda}{\Delta x}\,.
\end{equation}
Note that condition \eqref{frau} is indeed Fraunhofer's condition of
classical optics and Eq.~\eqref{I3} is the corresponding formula for
the intensity of classical Fraunhofer diffraction theory \cite{wolf},
according to which the large-distance intensity of a diffracted field
is the squared modulus of the Fourier transform of the field
distribution on the diffractive grating. (For example, for a
double-slit diffractive grating, with aperture $d$ of each slit,
distance $D$ between the slits, and with $\psi_0$ being the
characteristic function of the slits, Eq.~\eqref{I3} becomes the
standard formula for the intensity in the double-slits experiment,
namely,
\begin{equation}\label{two-slits}\nonumber
 \frac{I_0}{2}
   \:\mbox{sinc}^2\bigg(\frac{\pi d x}{\lambda L}\bigg)\, \bigg [1 +
\cos \bigg (\frac{2\pi Dx}{\lambda L}\bigg )\bigg ]\,,
\end{equation}
where $I_0$ is the intensity detected for $x=0$ and, as usual,
$\mbox{sinc}(x)\,\equiv x^{-1}\sin (x)$.)

In this regard, it should be observed that condition \eqref{condo}
leading to $I_3(x)$ is a known condition \cite{nino, 7steps}. It
corresponds to a ``large time'' regime
\begin{equation}\label{timetau}
\tf\gg\tau\equiv \frac{M(\Delta x)^2}{\hbar}= \frac{\Delta x}{\Delta p_x/M}
=\frac{\Delta x}{\Delta v_x}\end{equation}
for which the solution of Schr\"odinger's equation is approximated by
\begin{equation}\nonumber
\psi({x},t )\sim\sqrt{\frac{M}{i\hbar t}}\:
e^{i\frac{M { x}^2}{2\hbar t}}\:\widehat{\psi}_0\bigg
(\frac{M{x}}{\hbar t}\bigg )\,. \label{psiFT}
\end{equation}
Such a wave is what in \cite{7steps} has been named ``local
  plane wave'': a wave that locally looks like a packet, having
amplitude and local wave number that are slowly varying over
distances of the order of the local de Broglie wavelength. Such waves
are associated with classical motion of particles and a rough
estimate of the time needed for the formation of such waves is indeed
the time $\tau$ above \cite{7steps}. So, evidence to the contrary
notwithstanding, the particle motion in the Fraunhofer region, that
is, the particle motion on the time scale \eqref{timetau}, is indeed
classical motion.

It might be useful to compare the domain of validity of the various
approximations. Approximation $I_2(x)$ holds under the spatial
condition \eqref{sL}, that is on the time scale
\begin{equation}\nonumber
\tf \gg \frac{M\lambda }{2\pi \hbar} \Delta y\,.
\end{equation}
The temporal and spatial conditions leading to the Fraunhofer-like
approximation $I_3(x)$ are deduced by Eq.~\eqref{condo} for $t=\tf$.
For an initial packet with $\Delta x\sim~\Delta y$, comparison of
$\Delta x \ll L$ (leading to $I_2(x)$) with $\Delta x\ll~\sqrt{\lambda
  L}$ (leading to $I_3(x)$) shows that the Fraunhofer approximation is
realized on much larger space and time scales.

A final remark: It could be objected that it is indeed \eqref{I2} the
basic formula for the statistical predictions of detection
experiments---after all, it is this formula that seems to correspond
directly to the standard statistical interpretation of the wave
function.  {\em This objection, however, misses the point altogether:}
$I_2(x)$ is {\em only} an approximation; the time at which the
particle crosses the screen is typically random, and it can be treated
as the deterministic quantity given by the time of flight \emph{only}
when condition \eqref{sL} is satisfied.  In a sense, it is true that
in the regime of large distances there is no experimental difference
between $I(x)$ and the approximations we have considered (indeed, this
is a consequence of what it has been proven with great generality
in~\cite{nino2,det3}).  However, experimental research on near field
interferometry may explore regimes in which these approximations fail,
e.g., when statistical fluctuations in the arrival time become
experimentally relevant, as we shall comment in Section~\ref{newp}.
Hence the need of an \emph{exact} formula for the intensity. And,
while the standard quantum formalism fails to provide the exact
expression for the intensity, the formula given by Eq.~\eqref{curint},
which clearly {\em looks right}, is naturally deduced from first
principles of Bohmian mechanics~\cite{nino}.

\subsection{Interaction with the  environment}
\label{iefi2}

In the more general case of a quantum particle interacting with its
environment, the evolution cannot be any more treated in terms of
one-particle Schr\"odinger equation because entanglement can be fastly
developed. In this case, the statistical predictions concerning
experiments performed on the particle are governed by the {\em reduced
  density matrix} $\rho({\bf r}, {\bf r}')$, which is obtained from
the wave function describing the particle and its environment by
integrating out the configurational degrees of freedom of the
environment; for example, for an environment with N particles and
total wave function $\Psi({\bf r}, {\bf r}_{1},\ldots, {\bf r}_{N})$,
the reduced density matrix is given by
\begin{equation}
\rho({\bf r}, {\bf r}')=\!\!\int\! \ud{\bf r}_{1}\ldots \ud{\bf r}_{N}
\Psi({\bf r}, {\bf r}_{1},\ldots, {\bf r}_{N})\; \Psi({\bf r}', {\bf
  r}_{1},\ldots, {\bf r}_{N})^* \!\!.
  \label{reduceddm}
\end{equation}

As far as detection experiments are concerned, the following natural
generalization of Eq.~\eqref{basicequation} can be put forward: The
joint probability that the particle crosses the surface element $d{\bf
  S}$ of the surface $S$ at the point ${\bf r}$ in the time between
$t$ and $t+dt$ is still given by Eq.~\eqref{basicequation}, but with
current ${\bf J}$ given now by
\begin{equation}
\mathbf{J}({\bf r},t) =\frac{\hbar}{M}\,{\rm
    Im}\left[ \nabla_{{\bf r}}\rho({\bf r},{\bf r}',t) \right]_{{\bf
      r}'={\bf r}}.
\label{basicflux2}
\end{equation}
A detailed derivation of this result will be given elsewhere
\footnote{Here we just observe that the derivation of
  Eq.~\eqref{basicequation} given in~\cite{nino} can formally be
  extended to quantum states described by the reduced density matrix
  $\rho({\bf r},{\bf r}')$ by noting that: 1) the analysis
  in~\cite{nino} extends to ``{\em conditional wave functions}''; 2)
  the reduced density matrix $\rho({\bf r},{\bf r}')$ is indeed the
  projector onto the pure state defined by the conditional wave
  function {\em averaged} with respect to the ``{\em quantum
    equilibrium distribution}''. The notions of ``{conditional wave
    function}'' and ``{quantum equilibrium distribution}'' have been
  introduced in \cite{QE} and are crucial for a proper understanding
  of the empirical import of Bohmian mechanics. The only delicate
  point is current positivity. In order to ensure that the statistics
  of escape time and position is $\mathbf{J}$, current positivity
  should hold for all the conditional wave functions whose average
  leads to $\rho({\bf r},{\bf r}')$.}. In this regard, a key
observation is that ${\bf J}$ given by Eq.~\eqref{basicflux2} is
indeed the right probability current entering in the continuity
equation for the probability density of position $\rho({\bf r},{\bf
  r}, t)$,
\begin{equation}\nonumber
\frac{\partial \rho({\bf r},{\bf r}, t)}{\partial t} + \nabla_{\bf r}
\cdot {\bf J({\bf r},t)} = 0.
 \label{continuityequation2}
\end{equation}

We may now make more realistic the analysis of Section~\ref{sec2} by
allowing that, during the flight from the grating to the screen, the
particle of mass $M$ diffracted by the grating interacts with
particles of the environment (say, air molecules). As before, the
geometry is that of Fig.~\ref{fig1}, so that the dynamics is
effectively two dimensional, i.e., as before, ${\bf r} = (x,y)$.
Though the initial state produced by the grating needs not be anymore
a pure state, we maintain the physically reasonable assumption of
factorization at $t=0$,
\begin{equation}\nonumber
\rho({\bf r},{\bf r}',0) = \rho_{0}^{(x)}(x,x')\rho_{0}^{(y)}(y,y').
\label{fact}
\end{equation}
For an environment of N particles, the time evolution of $\rho({\bf
  r},{\bf r}',t)$ is that induced, according to Eq.~\eqref{reduceddm},
by the Schr\"odinger evolution of total wave function $\Psi=~\Psi({\bf
  r}, {\bf r}_{1},\ldots, {\bf r}_{N},t)$,
\begin{equation}\label{scevol}
i\hbar\frac{ \partial\Psi}{\partial t}= -\frac{\hbar^2}{2M}
{\nabla_{\bf r}}^2 \Psi + H_{0}^{\text{env}}\Psi +
 H_{\text{int}}\Psi,
\end{equation}
where $H_{0}^{\text{env}}$ is the total Hamiltonian of the $N$
particles (the sum of the kinetic and potential energies), and
$H_{\text{int}}$ is the interaction potential between the particle and
the other $N$ particles.

Accordingly, the exact formula for the probability density that the
particle crosses the screen at the point $x$ is given by
Eq.~\eqref{curint}, where $J_y$ is now the longitudinal component of
the probability current given by Eq.~\eqref{basicflux2}.  We shall now
simplify the expression for the intensity, in analogy with the
treatment of Section~\ref{sec2}, by exploiting the typical physical
conditions of interference experiments.

Motion along the $y$ direction is typically ``very fast,'' being
characterized by a ``very short'' wavelength $\lambda$, much smaller
than all the other relevant lengths scales (such as the spreads
$\Delta x$, $\Delta y$, and the screen distance $L$).  Accordingly,
we have an effective preservation of the factorization of the initial
state, and Eq.~\eqref{curint} becomes
$$
I(x)\sim \frac{\hbar}{M}\int^{+\infty}_{0}\!\!\!\!\!\!\!\!\ud t \
\rho^{(x)}(x,x,t)\ {\rm Im}\bigg [
\partial_{y}\rho^{(y)}(y,y',t)\vert_{y'=y} \bigg ]_{y=L}.
$$
Note that, due to the condition of fast motion along $y$ direction,
we may assume wave-packet motion, i.e., $\rho^{(y)}(y,y',t)=
\phi(y,t)\;\phi(y',t)^*$, and consider the evolution of the wave
packet $\phi$ to be classical. Thus, proceeding as in
Section~\ref{sec2} in going from Eq.~\eqref{curnum} to
Eq.~\eqref{I2}, and a part from a {\em caveat} we shall discuss
below, we arrive at the following approximation for the measured
intensity
\begin{equation}\label{Iint}
I_{2}(x) = \rho^{(x)}(x,x,\tf),
\end{equation}
where $\tf$ is, as before, the time of flight given by Eq.~\eqref{tf}.

The caveat is the following: in the free case a crucial condition for
the validity of Eq.~\eqref{I2} is that $\lambda\ll\Delta y$. In case
of environmental interaction the momentum spread increases due to
scattering events and the previous condition is no more sufficient to
assure a well defined de Broglie wavelength along the longitudinal
direction. An important consequence of the analysis of
Section~\ref{sec4} is that interaction with the environment produces
an effective reduction of the relevant length scales over which
quantum coherence is preserved and this reduction is controlled by
what we shall call the ``coherence length'' and denote by $\lco$. In
general, the validity of Eq.~\eqref{Iint} is assured by $\lambda \ll
\Delta y$ and $\lambda \ll \lco$; for relevant incoherence effects,
we have $\lco\lesssim\Delta y$ and thus the crucial condition
becomes: $\lambda\ll\lco$.

Let us consider the case of fullerene, and let the initial state be
the state at the moment of the splitting produced by the diffractive
grating.  It turns out that, at this time, the motion of fullerene
along the $y$-direction can be described by a narrow wave packet
translating with velocity $v$.  In fact, according to
Tab.~\ref{tab:table1}, the typical de Broglie wavelength for fullerene
is $\lambda\approx 10^{-12}$\,m. The analysis performed in Section
\ref{eec} leads to $\lco\approx 10^{-7}$~m (see
Tab.~\ref{tab:table3}), whence $\lambda\ll\lco$ (and since $L\approx
1\,$m, $\Delta y\ll L$). Thus, Eq.~\eqref{Iint} provides a good
approximation for the measured intensity of fullerenes.

{\bf N.B.} Eq.~\eqref{Iint} is the basic equation of this paper. In
order to avoid notational complexity, when no confusion will arise and
unless otherwise stated, we shall drop all the indices and simply
write $I(x)$ instead of $I_{2}(x)$ and $\rho(x,x,\tf)$ instead of
$\rho^{(x)}(x,x,\tf)$.

\section{Markovian Approximation}\label{subsec:JZ}
\label{section3}

In order to evaluate $I(x)$, we need to determine $\rho({\bf r},{\bf
  r}',t)$, the reduced density matrix at time $t$. In general, the
evolution of $\rho$ is highly non-Markovian, being the evolution
induced by Eq.~\eqref{scevol} via Eq.~\eqref{reduceddm}. For an
environment made of a gas at low pressure we may rely on the
Markovian approximation provided by the model of Joos and
Zeh~\cite{joze85}.  We shall now recall the {}essential ingredients
of this model and refer to literature for a thorough
discussion~\cite{giulini, flega90}.  (For some basic steps towards a
rigorous derivation see~\cite{figari}; in this regard see
also~\cite{andrea}; for a more general analysis of quantum Brownian
motion see~\cite{bassano, bassano2}).

This model aims at providing an autonomous evolution equation for an
heavy particle moving in a gas of light particles under the
approximation of negligible friction. To get a handle on the model,
consider a single collision of the heavy particle, of mass $M$, with
a light particle of the medium, of mass $m$. If $M\gg m$ the time
scale $\tau_s$ of a single-scattering process is short with respect
to the typical time scale $t$ of evolution of the heavy particle.
Thus, Born-Oppenheimer adiabatic approximation applies, and the
dynamics of the center of mass of the heavy particle can be
considered as \em frozen \em in the time $\tau_s$. Accordingly, if
$\Psi({\bf r})$ and $\chi({\bf r}_l)$ are, respectively, the wave
functions of the heavy particle and of the light particle before the
collision, the wave function of the composite system after the
collision is $\Psi({\bf
  r})\chi_{{\bf r}}({\bf r}_l)$, where $\chi_{{\bf r}}({\bf r}_l)$ is
the outgoing wave function of the light particle, scattered off at
the point ${\bf r}=(x,y,z)$. As a consequence, the final state of the
heavy particle is described by the reduced density matrix $\Psi({\bf
  r})\,\Psi({\bf r}')^*\:\langle \chi_{{\bf r}'}|\chi_{{\bf
    r}}\rangle$.

For arbitrary initial density matrix $\rho({\bf r},{\bf r}')$, and
many independent individual scattering events, the variation in the
time $\Delta t$ of $\rho({\bf r},{\bf r}')$ due to collisions is then
\[
\Delta \rho({\bf r},{\bf r}')\:\sim \: -\mathcal{N}\,\Delta t
\bigg(1-\overline{\langle{\chi_{{\bf r} '}}| {\chi_{{\bf
        r}}}\rangle}\:\bigg)\,\rho({\bf r} ,{\bf r}'),
\]
where $\mathcal{N}$ is the mean number of collisions per unit of time
and $\overline{\langle{\chi_{{\bf r}'}}| {\chi_{{\bf r}}}\rangle}$
denotes the average with respect to a suitable ensemble of light
particle wave functions. By taking into account also the rate of
change of $\rho$ due to the free dynamics, one arrives at the master
equation of Joos and Zeh:
\begin{equation}\label{din}
\frac{\partial \rho}{\partial t}= \mathcal{L}_0\,
\rho +\mathcal{L}_{\!I}\, \rho\,,
\end{equation}
where
$$
\mathcal{L}_0\, \rho =-\frac{i}{\hbar} [H_0,\rho] =
\frac{i\hbar}{2M} \left[ \nabla^2,\rho\right]
$$
and
\begin{equation}\label{JZ}
\big(\mathcal{L}_{\!I} \rho\big)({\bf r},{\bf r}')=\: -\mathcal{N}
\bigg(1-\overline{\langle{\chi_{{\bf r }'}}| {\chi_{{\bf r}
}}\rangle\!}\:\bigg)\,\rho({\bf r} ,{\bf r} ').
\end{equation}

\subsection{Estimation of environmental coupling}\label{eec}

\begin{table}[t]
\caption{\label{tab:table1}Physical parameters of fullerene experiments
\cite{JMO2000}.}
\begin{ruledtabular}
\begin{tabular}{ll}
Mass of fullerene $\fulle$:&$M\:\approx\:1.2\times 10^{-24}\,\mathrm{Kg}$\\
Radius of $\fulle:$ &$R\:\approx\:3.5\times 10^{-10}\,\mathrm{m}$ \\
Temperature of $\fulle$:&$\Theta_{F}\:\approx\:900\,\mathrm{K}$\\
Environmental temperature:&$\Theta_\mathcal{E}\:\approx\:300\,\mathrm{K}$\\
Mean wavelength \footnote{Mean values are deduced by the measured velocity
distribution characterizing the fullerene beam outgoing from the oven
(see Eq.~\eqref{supersonic} below).} of $\fulle$: &$\lambda\:\approx 2.5\times
10^{-12}\,\mathrm{m}$\\
Mean time of flight$^{a}$&$\tf \:\approx\:6\times 10^{-3}\,\mathrm{s}$\\
Grating--screen distance:& $L\:=\: 1.25 \,\mathrm{m}$\\
Collimator aperture: & $a\:=\: 10^{-5}\,\mathrm{m}$\\
Effective slits width: &$d\:\approx\: 3.6\times 10^{-8}\,\mathrm{m}$\\
Grating period: & $D\:=\: 10^{-7}\,\mathrm{m}$\\
\end{tabular}
\end{ruledtabular}
\end{table}

For a complete specification of the RHS of \eqref{din}, we need to
evaluate the interaction operator \eqref{JZ} related to the different
processes inducing entanglement of $C_{60}$ with surrounding
environment: scattering events (with thermal photons and air
molecules) and photon emission. Such an evaluation of the interaction
operator is rather standard, and can be found in the literature on the
Joos and Zeh model---modulo some numerical values that we have
corrected, and with the exception of our treatment of decoherence due
to photon emission that is slightly different from what can be found
in the literature (see, e.g.,~\cite{alicki, brukner}).

\subsubsection{\label{thermalphoton}Scattering with thermal photons}

In fullerene experiment, the environmental temperature is
$\Theta_\mathcal{E}\approx 300\,\mathrm{K}$ and thus the wavelength of
thermal photon is $\lambda_{ph}=hc/(\kb \Theta_{\mathcal{E}})\approx
4.8\times 10^{-5}\,\mathrm{m}$. As we shall see in Section
\ref{incoherence} (Eqs.~\eqref{geom-inco}--\eqref{xx'} and relative
evaluation in Tab.~\ref{tab:table3}), because of the incoherent
preparation of the beam, we have that $\vert x-x'\vert\lesssim
\ell_0\approx 10^{-7}$m. Under this condition, Eq.~\eqref{JZ} assumes
the form (see~\cite{joze85, giulini})
$$
\big(\mathcal{L}_{\!I} \rho\big)(x,x')\:=\:-\Lambda_{\rm ph}
^{\!(scat)}\vert\, x -x \,'\vert^2\rho(x,x')\,
$$
with
\begin{equation}\label{rey}
\Lambda_{\rm ph}^{\!(scat)}
\!\!=8!\,\frac{8\,c\,a^6}{3}\:\bigg\vert\frac{\epsilon_r -1
}{\epsilon_r +2}\bigg\vert^2\!\!\zeta(9)
\bigg(\frac{2\pi}{\lambda_{ph}}\bigg)^{\!9}\!\!\!\approx\,2.4\times
10^{2}\,\mathrm{m}^{-2}\mathrm{s}^{-1}\!\!,
\end{equation}
where the fullerene molecule has been modeled as a dielectric sphere
with the dielectric constant $\epsilon_r\approx 4$~\cite{epsfulle}
and $\zeta(9)\approx 1$, with $\zeta(z)$ representing the Riemann
$\zeta$~function.  The previous relation has been obtained in the
regime of Rayleigh scattering (since fullerene radius $R\approx
3.5\times 10^{-10}\, \mathrm{m}$ is much smaller than $\lambda_{ph}$)
and by using the Planck distribution for environmental photons
\footnote{Eq.~\eqref{rey} differs from the analog in~\cite{joze85}
  and~\cite{giulini} for numerical constants, which here have been
  corrected.}.

\subsubsection{\label{airmolecules}Scattering with air molecules.}

Air molecules, with a mean mass $m_{air}\!\approx\! 4.8\times
10^{-26}\,\mathrm{Kg}$, at the temperature $\Theta_\mathcal{E}\approx
300\,\mathrm{K}$ have a de Broglie wavelength $ \lambda_{air}=h
/\sqrt{2\pi m_{air}\kb
  \Theta_{\mathcal{E}}}\approx\,10^{-11}\,\mathrm{m}\, \ll\ell_0$ (see
Tab.~\ref{tab:table3}). Thus, assuming a Maxwell-Boltzmann
distribution for air molecule velocity, it follows from the analysis
performed in~\cite{flega90} (Eq.~(2.17)) that
\begin{equation}\label{lambda.aria}
\big(\mathcal{L}_{\!I} \rho\big)(x,x')\:=\:\left\{
\begin{array}{ll}\displaystyle \sigma_{\rm tot}\,P(\Theta_{\mathcal{E}})\,
\sqrt{\frac{32\pi\,}{\kb\, \Theta_{\mathcal{E}}\, m_{air}}}\:&\displaystyle
\text{for $x \,\neq\,x \,'$}\\[0.5cm]
0&\text{for $x \,=\,x \,'$,}\\[0.2cm]
\end{array}\right.
\end{equation}
where $P(\Theta_\mathcal{E})$ is the pressure at the temperature
$\Theta_\mathcal{E}$ and $\sigma_{\rm tot}$ is the total cross section
for scattering events.

In the case of fullerene experiment $P \approx 5\times
10^{-6}\,\mathrm{Pa}$ and $\sigma_{\rm tot}\approx
9\times10^{-18}\,\mathrm{m}^2$ \footnote{The values of pressure and
  cross section were kindly supplied by Dr. Olaf Nairz.}.  If $F_{\rm
  air}(\infty)$ is the constant value assumed by
$\big(\mathcal{L}_{\!I} \rho\big)(x,x')$ for $x\neq x'$, from
Eq.~\eqref{lambda.aria} we get $F_{\rm air}(\infty)\:\approx\:
32\,\mathrm{s^{-1}}$. In order to make a comparison between the
different decoherence sources involved in diffraction experiments, we
can introduce an effective localization factor $\Lambda$ also for air
scattering events. Given a pair of slits at the distance $nD$ in a
periodic grating of period $D$ we have
\begin{equation}\label{Lambdair}
\Lambda_{\rm air}(n)\:=\:\frac{F_{\rm air}(\infty)} {(n\,D)^2}\:.
\end{equation}
For adjacent slits ($n=1$) the localization factor $\Lambda_{\rm air}$
assumes its greatest value
\begin{equation}\label{L1air}
\Lambda_{\rm air}(n=1)\:\approx\:3.2\times
10^{15}\,\mathrm{m^{-2}\,s^{-1}}.
\end{equation}

\subsubsection{\label{secemiss}Photon emission}

The model of Joos and Zeh can be extended to the description of photon
emission processes. In fact, also in this case, the wave function of
the composite system after a single emission event is, in general,
$\Psi({\bf r})\chi_{{\bf r}}({\bf r}_l)$, where $\Psi({\bf r})$ is the
initial wave function of fullerene and $\chi_{{\bf r}}({\bf r}_l)$ is
the outgoing wave function of the photon emitted in ${\bf r}$. Since
emission time scale, in analogy with scattering events, is much faster
than characteristic time of fullerene free dynamics, the state
$|\chi_{{{\bf r} }}\rangle $, in position representation and
asymptotically in time, is well described by spherical waves
$$
\langle {\bf r}_l\,|\chi_{{{\bf r} }}\rangle\:\propto\:
\frac{e^{\,i\,k\,\vert\,{\bf r}_l-{\bf r} \,\vert}}{\vert \,{\bf
    r}_l-{\bf r} \,\vert}\,,
$$
where $k$ is the wave number of emitted photons. It follows that
$$
\langle{\chi_{{\bf r }'}}| {\chi_{{\bf r} }}\rangle\:=\: \frac{\sin
  (k\,\vert {\bf r} -{\bf r} \,'\vert)}{k\,\vert {\bf r} -{\bf r}
  \,'\vert}\:\equiv\:\mathrm{sinc}(k\,\vert{\bf r} -{\bf r} \,'\vert).
$$
According to Eq.~\eqref{JZ} and assuming that fullerene diffraction
can be effectively treated as a one-dimensional problem, the
interaction operator for photon emission becomes
\begin{equation}\label{sinc}
\big(\mathcal{L}_{\!I} \rho\big)(x,x')\:=\:
-\mathcal{N}\,\bigg[1-\overline{\mathrm{sinc}(k\,\vert x -x
\,'\vert)\!}\,\bigg] \,\rho(x ,x \,').
\end{equation}

In fullerene experiment there are essentially two channels for photon
emission: the black-body radiation and the disexcitation of internal
vibrational energy levels. In particular, for black-body radiation at
the fullerene temperature of $\Theta_{F} \approx 900\,\mathrm{K}$, we
have a mean wavelength of emitted photon equal to
$$
\lambda_{em}^{(bb)}\:\approx\:1.6\times 10^{-5}\,\mathrm{m}.
$$
For decays of internal energy levels it was measured a peaked
infrared spectrum with the shortest wavelength~\cite{kratschmer}
$$
\lambda_{em}^{(vib)}\:\approx\: 7\times 10^{-6}\,\mathrm{m}\,.
$$
In both cases it results (see Tab.~\ref{tab:table3})
$\lambda_{em}\gg\ell_0$.  This permits to simplify Eq.~\eqref{sinc} by
the expansion of the sinc-function in powers of $k\,\vert x -x
\,'\vert$. Keeping the first no-constant term, we straightly obtain
\begin{equation}\nonumber
\big(\mathcal{L}_{\!I} \rho\big)(x,x')\:=\:
-\Lambda_{em}\,\vert\,x -x \,'\vert^2\,\rho(x ,x \,'),
\end{equation}
with
\begin{equation}\label{Lambda_em}\nonumber
\Lambda_{\rm em} = \frac{\mathcal{N}\,\overline{k^2}}{6}
\end{equation}
(in agreement with what obtained by Alicki~\cite{alicki}).

In the following, we shall calculate the mean value $\overline{k^2}$
for the two different channels of photon
emission.\\

\paragraph{Black-body radiation.} The probability distribution of the
wave number $k$ is given by the Planck law
$$
n(k)\,\ud k\:=\:\varepsilon\,\bigg(\frac{\hbar\,c} {\kb
  \Theta_{F}}\bigg)^3\,\frac{1}{2\zeta (3)}\:\frac{k^2\,\ud k}
{e^{\frac{\hbar\,c}{\kb \Theta_{F}\!}\:\,k}-1},
$$
where $\varepsilon\approx 4.5\times 10^{-5}$ is the emissivity of
fullerene at $\Theta_{F}\approx 900$\,K~\cite{emit} and $\zeta
(3)\approx 1.2$.  Then Eq.~\eqref{Lambda_em} becomes
\begin{equation}\label{Lambdabb}
\Lambda_{\rm ph}^{\!(bb)}\: =\:\frac{\mathcal{N}}{6}\int \ud k
\:k^2\:n(k)\:=\:\frac{8\pi ^2\,\zeta(5)}{\zeta(3)}\:
\frac{\mathcal{N}\,\varepsilon}{(\lambda^{(bb)}_{em})^2}\,,
\end{equation}
with $\zeta(5)\:\approx\:1.04$.

The number of emitted photons per unit of time can be estimated as
$\mathcal{N}\sim E^{(bb)}/(\kb \Theta_{F})$, where $E^{(bb)}$ is the
total energy emitted per unit of time and $\kb\,\Theta_{F}\approx
0.08\,\mathrm{eV}$ represents the single photon energy. By integrating
the Planck distribution, one evaluates $E^{(bb)}\!\!=~\varepsilon S\sigma\Theta_{F}^4\approx~16\,\mathrm{eV} / \mathrm{s}$, where
$S=4\pi R^2$ is the total surface of fullerene macromolecules
($R\approx 3.5\times 10^{-10}\,\mathrm{m}$) and $\sigma$ is the
Stefan-Boltzmann constant. It results $\mathcal{N} \approx
200\,\mathrm{coll/sec}$, and thus
 \begin{equation}\label{L1bb}
 \Lambda_{\rm ph}^{\!(bb)} \:\approx\:
2.5\times 10^{9}\,\mathrm{m^{-2}\,s^{-1}}.
\end{equation}

\paragraph{Decay of internal vibrational energy levels.} Since we lack
of a model able to describe decays of internal energy levels for
fullerene, we directly refer to the results of experimental
spectroscopy. Since infrared spectrum shows a peaked structure, we can
write
\begin{equation}\label{Lambdavib}
\Lambda_{\rm ph}^{\!(vib)}\:\equiv\: \frac{\mathcal{N}\,
\overline{k^2}}{6}\:\lesssim\: \frac{\mathcal{N}\,
(k^{*})^{2}}{6} \:\approx \:5\times
10^{13}\,\mathrm{m}^{-2}\mathrm{s}^{-1}\:,
\end{equation}
where $k^*\approx 9\times 10^{5}\,\mathrm{m}$ is the wave number
related to the most energetic spectral line (see Fig.~4
in~\cite{kratschmer}) and $\mathcal{N}\approx 400\,
\mathrm{coll}/\mathrm{s}$~\cite{arnzei99}.\\

A direct comparison between evaluation \eqref{rey}, \eqref{L1air},
\eqref{L1bb} and \eqref{Lambdavib}, reported in
Tab.~\ref{tab:table2}, shows that the main decoherence processes are
scattering with air molecules (especially for adjacent slits, cf.
Eq.~\eqref{Lambdair}), followed by photon emission due to decay of
internal vibrational energy levels.

\begin{table}[h!]
\caption{\label{tab:table2}Sources of decoherence in the conditions
of fullerene experiments \cite{arnzei99,JMO2000}.}
\begin{ruledtabular}
\begin{tabular}{ll}
Decohering event &$\Lambda\ (\mathrm{m}^2\mathrm{s}^{-1})$\\
\hline\\[-4pt]
{Scattering}&\\
\hspace{10pt} with thermal photons:&$\Lambda_{\rm  ph}^{\!(scat)}\:
\approx\:2.4\times 10^{2},$ \\[4pt]
\hspace{10pt} with air molecules:&$\Lambda_{\mathrm air}\:
\lesssim\: 3.2\times 10^{15}.$\\[5pt]
Photon emission&\\
\hspace{10pt} black-body radiation:&$\Lambda_{\rm ph}^{\!(bb)}\:
\approx\: 2.5\times 10^9,$ \\[4pt]
\hspace{10pt} decay of excited states:&$\Lambda_{\rm  ph}^{\!(vib)}
\:\lesssim\: 5\times 10^{13}.$ \\[10pt]
Global effect: &$\Lambda \:\lesssim\: 3.3\times 10^{15}.$\\
\end{tabular}
\end{ruledtabular}
\end{table}

\subsection{The effective master equation}

According to the foregoing analysis, the Joos-Zeh equation \eqref{din}
effectively reduces to
$$
i\hbar \frac{\partial \rho}{\partial t}= [H_0,\rho] -i\Lambda
[x,[x,\rho]]
$$
or, more explicitly,
\begin{widetext}
\begin{equation}
i\hbar\frac{\partial \rho(x,x',t)}{\partial t}= \frac{\hbar^2}{2M} \left(
\frac{\partial^2}{\partial {x'}^2}- \frac{\partial^2}{\partial {x}^2}
\right) \rho(x,x',t)
-i \Lambda\:(x-x')^2\rho(x,x',t)
\label{onedim}
\end{equation}
with
\begin{equation}\nonumber
\Lambda\:\equiv\:\:\Lambda_{\rm
    air}\,+\,\Lambda_{\rm ph}\,,
\end{equation}
where $\Lambda_{\rm air}$ is given by Eq.~\eqref{Lambdair} and the
three terms in $ \Lambda_{\rm ph}\:\equiv\:\Lambda_{\rm
  ph}^{\!(scat)}\,+\,\Lambda_{\rm ph}^{\!(bb)}\,+ \,\Lambda_{\rm
  ph}^{\!(vib)}$ are given respectively by Eqs.~\eqref{rey},
\eqref{Lambdabb} and \eqref{Lambdavib}.

Eq.~\eqref{onedim} is a well-known equation and its solutions are
readily obtained (see, for example, App. 2 of Joos in~\cite{giulini})
\begin{equation}\label{boltzman-gen}
 \rho(x,x',t)\:=\:\iint \ud x_0 \:\ud
 x_0'\:\:K(x,x',t; x_0, x_0',0)\:\,\rho_{0}( x_0, x_0')\,,
\end{equation}
where
\begin{equation}\label{kgen}
K(x,x',t; x_0, x_0',0)=\frac{M}{2\pi\hbar t}\: \exp\bigg\{ \frac{i
M}{2\hbar\,t}\,\big[
(x- x_0)^2-(x'-{ x_0'})^2\big]\bigg\}\exp\bigg\{-\frac{\Lambda\,t}{3}\,
\bigg[(x-x')^2+( x_0- x_0')^2+(x-x')( x_0- x_0')\bigg]\bigg\}.
\end{equation}
Notice that the first exponential describes the free dynamics, while
the second takes into account the interaction with the external
environment.
\end{widetext}

\section{\label{incoherence}Preparation of the initial state}

In order to determine $\rho(x,x',t)$ and thereby evaluating the
intensity on the screen given by Eq.~\eqref{Iint}, we still need to
specify the initial density matrix $\rho_0(x,x')$, taking the initial
time $t=0$ at the moment of the splitting produced by the diffractive
grating.

Because of thermal production and in spite of the following
collimation, each fullerene wave function has a (mean) transversal
wave number $k_x$ (ideal collimation would correspond to $k_x=0$).
Thus, after the splitting, the macromolecule wave function is of the
form
\begin{equation}\label{inco}
\psi_0(x;k_x)\:=\:\bigg[\sum_{s}\varphi_{\!s}(x)\:\bigg]e^{i\,k_x\,x}\,,
\end{equation}
where $\varphi_s$ represents the $s$-th of the $N$ slit-shaped wave
packets outgoing from the grating.  The beam is an incoherent mixture
of such wave functions with wave number $k_x$ randomly distributed
according to a probability distribution $p\,(k_x)$. This distribution
depends on the geometry characterizing the collimation setup, which
reduces the wide thermally produced spread on $x$ direction. The
density matrix of the beam at $t=0$ is then
\begin{equation}\nonumber
\rho_0(x,x')=\int \ud k_x\ p\,(k_x)\ \psi_0(x;k_x)\;\psi_0(x';k_x)^*.
\end{equation}

\begin{figure}
\includegraphics [scale=0.45]{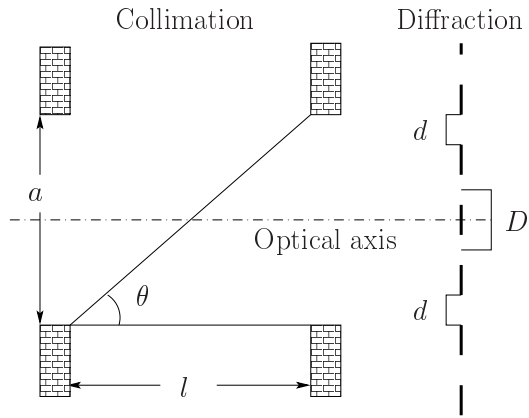}\caption{Collimation setup and diffraction grating.  For clarity the diagram is not in scale.}\label{fig2}
\end{figure}

Letting
\begin{equation}\label{rotilde}
\widetilde{\rho}_{0}(x,x')\,\equiv\,\sum_{s,s'}\varphi_{\!s}(x)\;
\varphi_{s'}(x')^*,
\end{equation}
we obtain
\begin{equation}\label{rotilda}
\begin{split}
  \rho_0(x,x') \,&=\:\widetilde{\rho}_{0}(x,x')\int \ud
  k_x\ p\,(k_x)\:e^{-i\,k_x\,(x'-x)}\\
  &=\:\sqrt{2\pi}\,\widetilde{\rho}_{0}(x,x')\:\widehat{p}\,(x'-x)\,,
\end{split}
\end{equation}
where $\widehat{p}$ is the Fourier transform of $p$ and
$\widetilde{\rho}_{0}(x,x')$ has the meaning of the density matrix in
the ideal case of perfect collimation. A typical diffraction setup
consists in a periodic grating of period $D$, which we consider placed
symmetrically with respect to the optical axis as illustrated in Fig.~
\ref{fig2}). In other words, we consider
\begin{equation}\label{pac}
\widetilde{\rho}_{0}(x,x')\,=\,\sum_{s,s'}\varphi\Big(x+s\frac{D}{2}\Big)\;
\varphi\Big(x'+s'\frac{D}{2}\Big)^*,
\end{equation}
where $s,s'=\pm 1,\pm 3,...,\pm (N-1)$ (for symmetry with respect to
the optical axis, $N$ is considered to be even). The size $\Delta x$
of the support of $\widetilde{\rho}$ is simply fixed by
\begin{equation}\label{nd}
\Delta x\sim ND\,.
\end{equation}
The general structure of Eq.~\eqref{rotilda} (for a treatment of
which we remind also to the section 9.1 of Joos in~\cite{giulini})
appears for any choice of the density matrix
$\widetilde{\rho}_0(x,x')$ and in every case in which a particle is
subjected to an uncontrollable source of random ``kicks'' which
produces instantaneous shifts in momentum, as in Eq.~\eqref{inco}.
Moreover, in case of random kicks with a mean momentum transfer
position-dependent (for example, in case of van der Waals interaction
between crossing particles and atoms of the grating), the effect on
the initial state consists in an effective reduction of the aperture
width \cite{fut}. A similar effect in molecular diffraction has been
already investigated in the framework of classical optics
\cite{grisenti}.

Now, in order to simplify the analysis, we shall adopt the convenient
and physically reasonable assumption of a Gaussian probability
distribution
\[
p\,(k_x)\:=\:\frac{1}{\sqrt{2\,\pi}\,\sigma_{k_x}}\,
\exp\bigg(-\,\frac{k_x^2}{2\,\sigma^2_{k_x}}\bigg )\,
\]
so that
\begin{equation}\label{geom-inco}
\rho_0(x,x')\:=\:\widetilde{\rho}_{0}(x,x')\:\exp
\bigg[-\frac{(x-x')^2}{2\ell^2_0}\bigg]\,,
\end{equation}
where we have defined
\begin{equation}\label{xx'}
\ell_0\,\equiv\, \sigma^{-1}_{k_x}.
\end{equation}

This quantity, which will play a relevant role in the following
analysis, will be called the {\em coherence length} (at time $t=0$).
We note that for $\ell_0\le\Delta x$ there is a bound on the length
on which the macromolecules can be coherent, expressed by $\ell_0$
itself. In particular, only beams characterized by an initial
coherence length $\ell_0\gtrsim D$ may produce a coherent
superposition of wave packets, and thus interference fringes, on the
detection screen. On the other hand, for $\ell_0 \gg \Delta x$, the
damping shown in Eq.~\eqref{geom-inco} does not take effect, and the
preparation of the initial state results to be coherent on the whole
support $\Delta x$.

Now we consider explicitly a typical diffraction experiment with
macromolecules (see Fig.~\ref{fig2}), in which the collimation
apparatus consists in two identical slits with aperture $a$, at a
distance $l\gg a$. The greatest drift velocity along the $x$
direction results to be $\vert v_x\vert _{\rm max}=v\theta$, where
$v$ is the macromolecule classical velocity along the optical axis
and $\theta=a/l$ is the angle under which a point situated in the
aperture of the first collimator sees the aperture of the second one
(since $l\gg a$, then the angle $\theta$ can be considered the same
for every point of the first collimator). Thus we can put
$3\sigma_{k_x}=Mv\theta/\hbar$ and so we obtain
\begin{equation} \label{valdiv}
\frac{1}{\ell^2_0}\,\equiv\,\sigma^2_{k_x}\:=\:\bigg(
\frac{Mv\theta}{3\hbar}\bigg)^2.
\end{equation}

An evaluation of the initial coherence length $\ell_0$ for fullerene
experiments is reported in Tab.~\ref{tab:table3}.

\section{\label{sec4}The Interference pattern}

Consider now the initial density matrix $\rho_0$ given by
Eq.~\eqref{geom-inco}.  Define
\begin{equation}\label{lcoh}
\frac{1}{2\ell(t)^{2}}\:\equiv\:\frac{\Lambda\,t}{3}\,+\,
\frac{1}{2\ell_0^2}\,,
\quad \mbox{i.e.,}\quad
\ell(t)\,\equiv\,\frac{\ell_0}{\sqrt{1+\frac{2\Lambda\,t}{3}\,\ell_0^2}}\,
\end{equation}
(note that $\ell(0)=\ell_0$). Then Eq.~\eqref{boltzman-gen} becomes
\begin{widetext}
\begin{equation}\nonumber
\rho(x,x',t)=\frac{M}{2\pi\hbar t}\!\iint \!\!\ud x_0\,\ud
 x_0'\:\exp\bigg\{\: \frac{iM}{2\hbar t}
\bigg[(x- x_0)^2-(x'- x_0')^2\bigg]-\:\frac{( x_0- x_0')^2}{2\ell(t)^{2}}
-\frac{\Lambda\,t}{3}\bigg[(x-x')^2+(x-x')( x_0- x_0')\bigg]\!\bigg\}
\widetilde{\rho}_0( x_0, x_0'),
\end{equation}
whence, from relation~\eqref{Iint}, the intensity on the screen is
given by
\begin{equation}\label{ro}
I_2(x)\ \equiv\ \rho(x,x,T)\ =\ \frac{M}{2\pi\hbar T}\!\iint\!\!\ud x_0\,
\ud  x_0' \exp \bigg\{\frac{iM}{2\hbar
\tf }\bigg[ x_0^2- x_0'^2+2x( x_0'- x_0)\bigg]-\frac{( x_0- x_0')^2}{2\lcoq}
\bigg\} \,\widetilde{\rho}_0( x_0, x_0'),
\end{equation}
\end{widetext}
where $x_0$ and $x_0'$ run along the slits crossed by the initial wave
function whose support is $\Delta x$ and $\lco\equiv\lco (\tf)$ is the
coherence length at the time of flight $t=\tf$.

As already argued for $\ell_0$, if $\lco \lesssim \Delta x$ the
exponential $\smash{\exp[-(x_0-x_0')^2/(2\lcoq)]}$ reduces from
$\Delta x$ to $\lco$ the length scale on which the initial state is
coherent. This scale is fixed from both geometry of the experimental
setup, i.e., the collimation apparatus and the distance $L$ between
grating and screen, and the physical conditions under which the
interferometry takes place, i.e., the momentum of the macromolecule
and the effect of the environment (see Eqs.~\eqref{lcoh}
and~\eqref{tf}). On the other hand, the above exponential does not
give any relevant contribution if $\lco~\gg~\Delta x$, and, from
Eqs.~\eqref{valdiv} and~\eqref{lcoh}, it follows that this occurs when
there are both good collimation ($\theta\approx0$) and negligible
coupling with surrounding environment ($\Lambda \approx 0$). In this
case we fall back to the treatment of section \ref{sec2}.

Note that interference fringes appear on the detection screen if
$\Delta x \gtrsim D$ and $\lco\gtrsim D$, i.e., if the molecules are
coherent at least on two contiguous slits. The numerical estimates of
$\ell_0$ and $\lco$ in the condition of fullerene experiments are
shown in Tab.~\ref{tab:table3}. In particular, note that $\lco\approx
D$ and thus interference is mainly due to adjacent slits. Moreover, a
comparison between the values of $\ell_0$ and $\lco$ shows that the
main mechanism which yields a loss of coherence is the angular
divergence of the beam \footnote{A recent proposal could lead to
  localization factors smaller by a factor of $2\pi$~\cite{sipe}.  In
  this case decoherence effects, which are already quite less relevant
  with respect the loss of coherence due to beam preparation, should
  be further negligible.}.

\begin{table}[h!]
\caption{\label{tab:table3} Comparison between losses of coherence
in fullerene experiments \cite{arnzei99,JMO2000}.}
\begin{ruledtabular}
\begin{tabular}{ll}
Initial coherence length ($t=0$):& $\ell_0\,\approx\,1.3\times 10^{-7}\,
\mathrm{m}$\\[1pt]
Coherence length at $t=\tf $:& $\lco\,\sim\, D\,=10^{-7}\,
\mathrm{m}$\\[1pt]
\end{tabular}
\end{ruledtabular}
\end{table}

\subsection{Fraunhofer approximation for the intensity}

We shall now proceed to an approximate evaluation of $I(x)$, relying
on conditions that are reasonable in common interferometry
experiments performed in far-field approximation (see App.~\ref{Y}
for a more refined evaluation in the case of a pair of Gaussian
shaped slits).

Note that
\begin{equation}\nonumber
\exp\bigg[\frac{iM( x_0^2- x_0'^2)}{2\hbar \tf} \bigg]\approx 1
\end{equation}
when
\begin{equation}\nonumber
\frac{M(x_0^2-x_0'^2)}{2\hbar \tf}=\frac{M(x_0+x_0')(x_0-x_0')}{2\hbar
  \tf}\ll 1
\end{equation}
and this condition is clearly satisfied in the Fraunhofer regime
\eqref{timetau}. Nevertheless, in the presence of a coherence length
$\lco \lesssim \Delta x$, we have relevant contributions in
integration \eqref{ro} just for $(x_0-x_0')\lesssim \lco$. Thus,
under the condition
\begin{equation}\label{condo2}
\frac{M \Delta x\,\lco}{\hbar \tf}
\ll 1\,,
\end{equation}
$I_2(x)$ gets approximated by
\begin{equation}\label{iI3}
 I_3 (x) =\frac{M}{2\pi\hbar T}\!\iint\!\! \ud x_0\, \ud x_0'
\exp \bigg[ \frac{iM x}{\hbar \tf} (x_0'-x_0)  \bigg]\,
 {\rho}_1(x_0,x_0'),
\end{equation}
where
\begin{equation}\nonumber
  {\rho}_1(x_0,x_0') =
\exp \bigg[-\frac{(x_0-x_0')^2}{2\lcoq}\bigg] \,
\widetilde{\rho}_0(x_0,x_0')\,.
\end{equation}
Introducing the Fourier transform of $\rho_1$
\[
\widehat{\rho}_1(k_0,k_0') = (2\pi)^{-1}\!\iint\! \ud x_0 \ud x_0'\,
e^{-i(k_0\,x_0 +k_0'\,x_0')}\:\rho_1 (x_0,x_0')\,
\]
we have
\begin{equation}\label{roBIS3}
I_3(x)= \frac{M}{\hbar T}\,  \widehat{\rho}_1(\bar{k}, -\bar{k}),
\quad\mbox{where}\quad\bar{k}\equiv\frac{M x}{\hbar \tf}=
\frac{2\pi x}{\lambda L}.
\end{equation}
This result is very analogous to Eq.~\eqref{I3} of section \ref{sec2}
with \eqref{condo2} replacing \eqref{condo} whenever
$\lco\lesssim\Delta x$. Also in this case \eqref{condo2} should be
rewritten in terms of the physical variables under control (cf.
\eqref{frau}), namely as
\begin{equation}\nonumber
 \frac{\lco}{L}\ll \frac{\lambda}{\Delta x}\,.
\end{equation}
This notwithstanding, there are some basic difference that should be
underlined: First, $\rho_1$ is not the initial state, but it is an
effective state that takes into account incoherence due to preparation
and to the time evolution. In fact $\lco$ depends on the physical and
geometrical variables of the experiment in the phase of preparation
and in its future development and it is progressively reduced by
increasing the time of flight. Second, unlike what typically
happens in the framework of classical optics and the theory of
scattering, it is no more useful to evaluate $\rho_1$ asymptotically
in time, since environmental-induced decoherence completely destroys
interference fringes at times too large.

In fullerene experiments, $\lco\approx 10^{-7}$~m $<\Delta x\approx
10^{-6}\,\mathrm{m}$ (for an estimate of $\Delta x$, see
Section~\ref{s5}). In this case the left-hand side (LHS) of
\eqref{condo2} is not at all negligible with respect to unity.
Anyhow, a more precise inspection of integration \eqref{ro}, with
$\widetilde{\rho}_0$ given by \eqref{pac}, shows that condition
\eqref{condo2} is a too strong demand and that approximation
\eqref{iI3} can be reasonably applied. So doing, the error made is
not completely negligible only for the pair of adjacent slits
farthest with respect to the optical axis. This error, however,
affects negligibly the sum involving the contributions of all the
slits.

Let us now compute $\widehat{\rho}_1(\bar{k}, -\bar{k})$ for
$\widetilde{\rho}_0$ expressed by \eqref{pac}, i.e., for the split of
macromolecules on a periodic grating of period $D$.  First we make
the change of variables $\xi\equiv x_0+sD/2$ and $\xi'\equiv
x_0'+s'D/2$, which leads to
\begin{widetext}
\begin{equation}\label{change}
\widehat{\rho}_1(\bar{k}, -\bar{k})=
\frac{1}{2\pi} \sum_{s,s'}e^{\,i\bar{k}(s
-s')\frac{D}{2}}\,e^{-\frac{[(s-s') D/2]^2}{2\lcoq}}
\iint \ud \xi \ud\xi'\,e^{-i \bar{k}\xi}\, e^{i
\bar{k}\xi'}\,e^{\frac{1}{2\lcoq}\big[(\xi'-\xi)(s'-s)D-(\xi'-\xi)^2\big]}\,
\varphi(\xi)\;\varphi(\xi')^*\,,
\end{equation}
and second we perform the Taylor series expansion of the real
exponential in the previous integral in the variable
$(\xi'-\xi)/\ell$ and about the point $\xi'=\xi$
\begin{equation}\label{bigapp}
\exp\bigg[\frac{(\xi'-\xi)(s'-s)D-(\xi'-\xi)^2}{2\lcoq}\bigg] = 1 +
\frac{(\xi'-\xi)(s'-s)D}{2\lcoq}
 + O\bigg[\bigg(\frac{\xi'-\xi}{\lco}\bigg)^2\bigg].
\end{equation}
\end{widetext}
The solution of Eq.~\eqref{change} is particularly handy whenever the
effects due to incoherence are negligible on a length scale of the
order of the slit width $d$ or, in other words, whenever the strength
of incoherence does not spatially resolve the single slit. This is
assured by a slit width much less than the coherence length
\begin{equation}\label{risdeco}
d/\lco\ll1.
\end{equation}
Under condition \eqref{risdeco} the LHS of Eq.~\eqref{bigapp} is well
approximated by $1$. In fact, since $\xi$ and $\xi'$ run within the
slit width $d$ and $(s'-s)D/2\lesssim \lco$ due to the damping
exponential $\exp\Big\{\!\!-\![(s-s') D/2]^2/(2\lcoq)\Big\}$ in Eq.
\eqref{change}, we have
\[
\frac{(\xi'-\xi)(s'-s)D}{2\lcoq}\ \lesssim\ d/\lco\ll 1.
\]
Thus the Fourier transform~\eqref{change} becomes
\begin{widetext}
\begin{equation}\nonumber
\widehat{\rho}_1(\bar{k}, -\bar{k})\:\sim\:\frac{1}{2\pi}
\sum_{s,s'}e^{\,i\bar{k}(s
  -s')\frac{D}{2}}\,e^{-\frac{[(s-s')
    D/2]^2}{2\lcoq}} \iint \ud \xi \ud\xi'\,e^{i
  \bar{k}\xi}\,e^{-i\bar{k}\xi}\varphi(\xi)\;\varphi(\xi')^*.
\end{equation}
\end{widetext}
(Although the rough condition \eqref{risdeco} is not directly
satisfied in fullerene experiments, the zero-order approximation of
\eqref{bigapp} can be reasonably applied in integration
\eqref{change}; for an evaluation of the error introduced the reader
can see App.~\ref{aQ}).

In the light of these considerations, the intensity pattern is well
approximated by
\[
I_4(x)\:=\:\frac{M}{\hbar \tf
  }\big\vert\widehat{\varphi}(\bar{k})\big\vert ^2\sum_{s,s'}
e^{\,\frac{i\bar{k}}{2 }(s-s') D}\,e^{-\frac{[(s' -s)
    D/2]^2}{2\lcoq}}\,,
\]
where $\widehat{\varphi}$ is the Fourier transform of $\varphi$.

Note that the sum of the terms with $s=s'$ simply gives $N$, while the
sum of the terms with $s \ne s'$ gives
$$
2 \sum_{s >s'}e^{-\frac{[(s' -s) D/2]^2}{2\lcoq}}\!\cos\bigg
[\frac{\bar{k}(s'-s)}{2}\bigg].
$$

By adding these two contributions and with $\bar{k}=~2\pi x/(\lambda
L)$ (see Eq.~\eqref{roBIS3}), we arrive at the suggestive
``Fraunhofer-like'' expression
\begin{widetext}
\begin{equation}\label{N}
I_4(x)\:\equiv\:I(x)\:=\:\frac{2\pi}{\lambda L}\,
\bigg\vert\,\widehat{\varphi}\bigg(
\frac{2\pi x}{\lambda L}
\bigg)\bigg\vert ^2 \,\bigg[
N+2\sum_{n=1}^{N-1}(N-n)\,e^{-\frac{(nD)^2}{2\lcoq}}\!\cos
\bigg(\frac{2\pi nD x}{\lambda L}\bigg)\bigg]
\end{equation}
\end{widetext}
(being understood that for $N=1$ the sum is zero), where $n=(s' -s)/2$
\footnote{Eq.~\eqref{N} holds under the conditions \eqref{condo2},
  $\tf \gg \tau$ with $\tau=M\Delta x\lco/\hbar$, and \eqref{risdeco},
  $d/\lco\ll 1$. These conditions imply a large superposition of the
  wave packets on the screen. In fact, the ratio $\mathcal{R}$ between
  the separation of the most distance slits ($\sim \Delta x$) and the
  size of the pattern $I(x)$ ($\sim \lambda L /2\pi d = \hbar \tf
  /Md$) becomes $\mathcal{R}\ll d/\lco$.}.

Equation \eqref{N} shows that, whereas all the $N$ wave packets
outgoing from the grating contribute to the intensity revealed on the
screen, the pairs of slits which concretely contribute to
interference oscillations are distant at most of the order of $\lco$,
due to the damping exponential in the sum. It follows that, for a
finite $\lco$, the interference pattern shows ``distortions'' in
fringe structure due to partially random preparation and decoherence,
but, being incoherent effects typically negligible on single-slit
space scale, fringe pattern is just modulated by the single-slit
diffraction profile, $\vert\widehat{\varphi}\vert ^2$, according to
classical optics Fraunhofer diffraction.

As already sketched before, it should be observed that the intensity
on the screen may show interference fringes only if the coherence
length is at least as long as the grating period, i.e.,
\begin{equation}\label{condint}
\ell(t)\gtrsim D
\end{equation}
(note that this inequality should be satisfied at least at the initial
time $t=0$). For positive times, recalling \eqref{lcoh}, we obtain
\begin{equation}\nonumber
t\lesssim\tau_c\equiv\frac{3}{2 \Lambda D^2}\bigg[1-\bigg(\frac{D}{\ell_0}
\bigg)^2\bigg],
\end{equation}
which provides an upper bound for the time of flight, i.e., an
evaluation for the \emph{effective coherence time} $\tau_c$. Note that
for the fullerene experiment, according to Tab.~\ref{tab:table1} and
Tab.~\ref{tab:table2}, it results
\[
\tau_c=4\times 10^{-2}\mbox{s}\,,
\]
which is indeed several times the value of the time of flight in this
experiment \footnote{Note that condition \eqref{risdeco}, $\lco\gg d$,
  is obviously consistent with \eqref{condint}, $\lco\ge D$, being
  $d<D$. In particular, for fullerene experiment, condition
  \eqref{risdeco}, via definition \eqref{lcoh}, leads to $\tf \ll
  \frac{3}{2\Lambda d^2} [1-(d/\ell_0)^2 ]\approx 0.3\,\mathrm{s}$.}.

The effective coherence time $\tau_c$ is clearly an upper bound for
the time of flight $T$, since interference fringes are detectable {\em
  only} within $\tau_{c}$. This shows that the ``geometrical optics''
limit in presence of decoherence requires more care than in the free
case. In particular, it can not be based on the standard
time-independent methods and the ``$t\to\infty$'' limit.

\begin{widetext}
\subsection{Extension to a generic angular divergence of the beam}

This section is devoted to generalize Eq.~\eqref{N} for a generic
transversal wave number probability distribution $p\,(k_x)$.
Introducing Eqs.~\eqref{kgen} and \eqref{rotilda} in
Eq.~\eqref{boltzman-gen}, the long-time asymptotic behavior of the
intensity becomes
$$
I_3(x)\:=\:\frac{M}{\sqrt{2\pi}\hbar \tf }\!\iint\!\! \ud x_{0}\,
\ud x_{0}' e^{i\bar{k}\,(x_{0}'-x_{0})-\frac{\Lambda \tf
    }{3}\,(x_{0}'-x_{0})^2} \,\widehat{p}\,(x_{0}'-x_{0})\,
\widetilde{\rho}_0(x_{0},x_{0}').
$$

By the same variable change which leads to Eq.~\eqref{change} and
making explicit $\widetilde{\rho}_0$ for a grating of period $D$ (cf.
Eq.~\eqref{pac}), we obtain
\begin{equation}\label{ext}
\begin{split}
  I_3(x)\:=\:&\frac{M}{\sqrt{2\pi}\hbar \tf
    }\sum_{s,s'}e^{\,i\bar{k}(s -s')\frac{D}{2}} \, e^{-\frac{\Lambda
      \tf }{3}[(s'-s) D/2]^2}\\
  &\times \iint \ud \xi \ud\, \xi'\, e^{\,-i\bar{k} \xi}\,
  e^{\,i\bar{k} \xi'} \, e^{-\frac{\Lambda \tf
      }{3}\big[(\xi'-\xi)^2+(\xi'-\xi)(s '-s)D\big]}\,
  \widehat{p}\,\big[\xi'-\xi+(s-s')D/2\big]\, \varphi(\xi)\;
  \varphi(\xi')^*.
\end{split}
\end{equation}
As discussed in the preceding section, note that $\vert
\xi'-\xi\vert\le d$ and that $\vert (s'-s)D/2\vert_{\rm max}\sim
(\Lambda \tf )^{-1/2}$ due to the damping term $e^{-\frac{\Lambda \tf
    }{3}[(s'-s) D/2]^2}$ in Eq.~\eqref{ext}. Thus, in case of
decoherence negligible on the single-slit length scale, i.e., for
$d(\Lambda \tf )^{1/2} \ll 1$, and for a slowly varying function
$\widehat{p}$, such that $\widehat{p}\,\big[\pm d+(s-s')D/2\big]\sim
\widehat{p}\,\big[(s'-s)D/2\big]\ \forall\,s,s'$, Eq.~\eqref{ext}
becomes

\begin{equation}\label{Ngen}
I(x)\:=\:
\frac{(2\pi)^{3/2}}{\lambda L}
\,\widehat{p}\,(0)\,
\bigg\vert\,\widehat{\varphi}\bigg(
\frac{2\pi x}{\lambda L}
\bigg)\bigg\vert ^2 \,\bigg \{
N+2\sum_{n=1}^{N-1}(N-n)\,e^{-\frac{\Lambda \tf }{3} (nD)^2}\,\bigg[
\frac{\widehat{p}\,(nD)}{\widehat{p}\,(0)}\bigg]\,\cos
\bigg(n
\frac{2\pi D x}{\lambda L}
\bigg)\bigg \},
\end{equation}
where $n=(s' -s)/2$ and $\bar{k}=2\pi x/(\lambda L)$. Note that the
assumption of slow variation of $\widehat{p}$ is directly assured by
a sufficient sharpness of the wave number distribution $p\,(k_x)$,
i.e.,
by $\Delta k_x \ll d^{-1}$.\\

\end{widetext}

\section{Quantum interferometry and classical diffraction theory}

\subsection{Comparison with geometrical optics}\label{subsec4}

In case of complete coherence, i.e., for $p\,(k_x)=\delta(k_x)$ and
$\Lambda =0$, Eq.~\eqref{Ngen} reduces to the well-known Fraunhofer
relation for optical diffractive patterns
\[
\begin{split}
  I(x) &\:=\: \frac{2\pi}{\lambda L} \,
  \bigg\vert\widehat{\varphi}\bigg( \frac{2\pi x}{\lambda L} \bigg) \,
  \sum_{n=0}^{N-1}e^{\,in\frac{2\pi Dx}{\lambda L}}
  \bigg\vert ^2\\
  &\ =\ \frac{2\pi}{\lambda L}\,
  \bigg\vert\widehat{\varphi}\bigg(\frac{2\pi x}{\lambda L}\bigg)
  \bigg\vert^2 \,\bigg [\frac{\sin(\pi NDx/\lambda L)}{\sin(\pi
    Dx/\lambda L)}\bigg ]^2.
\end{split}
\]

Moreover, for just two slits ($N=2$), i.e., for Young double-slit
interference, Eq.~\eqref{Ngen} becomes
\begin{equation}\nonumber
\begin{split}
  I(x)\:=\:&\frac{4\pi}{\lambda L}\,\widehat{p}\,(0)\,
  \bigg\vert\widehat{\varphi}\bigg(\frac{2\pi x}{\lambda L}\bigg)
  \bigg\vert
  ^2\\
  &\times\bigg [1 + \frac{\widehat{p}\,(nD)}{\widehat{p}\,(0)}
  \,e^{-\frac{\Lambda M\lambda L}{6\pi\hbar} D^2}\!\cos \bigg
  (\frac{2\pi Dx}{\lambda L}\bigg )\bigg ].
\end{split}
\end{equation}
This expression is very similar to that used in classical optics to
describe interference patterns due to partially coherent
electromagnetic fields~\cite{wolf}. In particular, note that the
\emph{damping term} for quantum interference oscillations
\begin{equation}\label{dt}
\mathcal{V_\mathrm{QM}}\ =\ \frac{\widehat{p}\,(nD)}{\widehat{p}\,(0)}\,
e^{-\frac{\Lambda M\lambda L}{6\pi\hbar} D^2}
\end{equation}
is the quantum-mechanical counterpart of the \emph{fringe visibility}
$\mathcal{V_\mathrm{CO}}$ of classical optics
\begin{equation}\label{conf}
\mathcal{V_\mathrm{QM}}\ \longleftrightarrow \ \mathcal{V_\mathrm{CO}}.
\end{equation}
Both for quantum and classical interferometry, the visibility
$\mathcal{V}$ is a measure of the distinctness of the fringes and is
defined by
\begin{equation}\label{vis}
\mathcal{V}\ =\ \frac{I_{\rm max}-I_{\rm min}}{I_{\rm max}+I_{\rm
min}}\,.
\end{equation}
The intensities $I_{\rm max}$ and $I_{\rm min}$ are, respectively, the
maximum and the minimum revealed on the detection screen in the
immediate neighborhood of the optical axis.

Now it is useful to recall an important result from the classical
theory of partial coherence. The pattern visibility
$\mathcal{V_\mathrm{CO}}$ of a quasimonochromatic field, equally
split by a pair of slits, coincides with the modulus of the
\emph{spectral degree of coherence} $\mu(\lambda)$ \cite{wolf, otto},
which characterizes the field correlation in the space-frequency
domain
\begin{equation}\label{ug}
\mathcal{V_\mathrm{CO}}\ =\ \vert\mu(\lambda)\vert.
\end{equation}
From Eqs.~\eqref{vis} and \eqref{ug} it follows that the degree of
spectral coherence is upper bounded by unit, value assumed in
condition of complete coherence (e.g., in case of laser radiation
diffraction).

According to the correspondence \eqref{conf}, the results of the
classical theory of partial coherence extends to quantum systems,
\emph{mutata mutandis}. For instance, in Section~\ref{simulations} we
shall show some interesting analogies concerning with \emph{temporal
  and spatial coherence} of beams, while in the following we underline
the differences existing between classical optics and quantum
mechanics.

First of all, the degree of coherence of quantum particles depends
both on the collimation of the macromolecular beam and on the
strength of interaction with the surrounding environment during the
time of flight (cf. Eq.~\eqref{dt}).  In optics, instead, the degree
of coherence of quasimonochromatic fields is only due to source
details. More particularly, the corruption of visibility of
interfering fields increases with the spatial extension of the
source, composed by a statistical ensemble of many independent
elementary radiators.

Moreover, in classical optics the explicit form of the degree of
coherence $\mu(\lambda)$ depends on the geometrical shape of the
source, while the damping term $\mathcal{V}_\mathrm{QM}$ depends both
on the features of the evolution kernel \eqref{kgen}, characterizing
the decoherence model, and on the geometrical details of the
collimation apparatus (cf. Eq.~\eqref{dt}).

\subsection{Fresnel regime and Talbot interferometry}

We would like now to comment on interferometry in the near-field
zone~\cite{klauser, chapman, nowak, kimble}, which has been recently
realized by means of $C_{70}$ beams~\cite{ztalbot, ztalbot2}.  Such
experiments show that, at distances from the diffraction grating
multiple of the length $2 D^2/\lambda$, images of the grating itself
are reconstructed (see also the optical Talbot effect~\cite{talbot,
  winthrop}).

Thus, by shifting another identical grating, placed behind the
previous one at a distance $2 D^2/\lambda$, the integrated signal
outgoing from the gratings periodically changes from its minimum (half
period displacement of the two gratings) to its maximum (complete
alignment).

If the influence of the environment is negligible, a treatment of
this effect in the spirit of Section ~\ref{sec2} can be performed. In
fact, given the correspondence between Helmholtz and stationary
Schr\"odinger equation, one can directly exploit the standard optical
techniques, such as Fresnel-Kirchhoff diffraction integrals in
Fresnel zone, with suitable boundary conditions---``transmission
functions''---at the gratings.  Indeed, this is what it has been done
(see, e.g.,~\cite{kimble, ztalbot3}) by means of the so-called
``paraxial approximation," assuming both gratings distances large
with respect to the grating period and an infinite number of slits.

In experiments with large molecules~\cite{ztalbot2}, it has been
observed that the visibility of the signal is progressively reduced
by increasing the pressure of environmental gas, a clear sign of
environmental {\em quantum} decoherence.  A quantitative explanation
of this effect---using the model of Joos and Zeh in order to suitably
modify the classical Fresnel-Kirchhoff description recalled
above---has been already provided in~\cite{ztalbot2}. A more
self-contained and thorough analysis, based on Eq.~\eqref{Iint}, will
be presented elsewhere~\cite{fut}. Here we shall provide just a
sketchy outline, referring to a theoretical treatment already present
in literature~\cite{schleich}.  In this last work, the form of the
propagator describing the free evolution of a quantum wave is shown,
split by a diffraction grating with a formally infinite number of
slits, i.e., with an associated initial density matrix of the form
\begin{equation}\label{rotalb}
\rho_{0}(x,x')\,=\!\sum_{j,j'\in \mathbb{Z}}\!\varphi(x+j
D)\;\varphi(x'+j' D)^*
\end{equation}
(note that, with respect to initial state expressed by~\eqref{rotilda}
and \eqref{pac}, here it is assumed perfect collimation and an
infinite number of slits). Let
$\smash{K^{\text{\scriptsize(free)}}_T}$ be the so called \emph{Talbot
  propagator}, including the sum over $j$ and $j'$ of \eqref{rotalb}
and, accordingly, providing the intensity pattern $I(x)$ in term of
the \emph{single} wave packet $\varphi(x)$:
\begin{equation}\label{int_tal}
I(x) \propto \int \ud x_0 \, \ud x_0'\:K^{\text{\scriptsize(free)}}_T
(x,x,t; x_0, x_0', 0) \,\varphi(x_0)\;\varphi(x_0')^*\,.
\end{equation}
It results (see e.g. \cite{schleich}) that, at a distance $\LT$ from
the grating equal to $2 D^2/\lambda$, or multiple of it (and
consequently at times $\tT$ multiples of $\LT/v=2MD^2/h$),
$\smash{K^{\text{\scriptsize(free)}}_T}$ reduces to

\begin{widetext}
\begin{equation}\label{Kta}
\begin{array}{ll}
\displaystyle K^{\text{\scriptsize(free)}}_T(x,x,\tT; x_0, x_0',0)&
\displaystyle=\frac{1}{D^2}
\sum_{j=-\infty}^{+\infty}\exp\bigg(-2\pi i j\frac{x-x_0}{D}\bigg)
\sum_{j'=-\infty}^{+\infty}\exp\bigg(-2\pi i j'\frac{x-x'_0}{D}\bigg)\\
&\displaystyle = \sum_{j=-\infty}^{+\infty}\delta(x-x_0+jD)\,\sum_{j'=
-\infty}^{+\infty}\delta(x-x'_0+j'D).
\end{array}
\end{equation}
Clearly, from this relation and Eq.~\eqref{int_tal} it is immediate to
verify that the final state is an exact reconstruction of the initial
one~\eqref{rotalb}.

By following the same steps leading to Eq.~\eqref{Kta}, but now using
the propagator~\eqref{kgen} which embodies the incoherence effects,
the Talbot propagator becomes
\[
\begin{array}{ll}
\displaystyle K^{\text{\scriptsize(env)}}_T(x,x,\tT; x_0, x_0',0)=
&\displaystyle \frac{1}{D^2}
\sum_{j=-\infty}^{+\infty}\exp\bigg(-2\pi i j\frac{x-x_0}{D}\bigg)\:
\exp\bigg\{- 2 j^2 \bigg[\frac{D}{\ell(\tT)}\bigg]^2\bigg\}\\
&\ \times\displaystyle \sum_{j'=-\infty}^{+\infty}\exp\bigg(-2\pi i
j'\frac{x-x'_0}
{D}\bigg)\:
\exp\bigg\{- (2 {j'}^2 + 4 j j')\bigg[\frac{D}{\ell(\tT)}
\bigg]^2\bigg\}\,,
\end{array}
\]
where $\ell(\tT)$ is the coherence length computed at the Talbot time
$\tT$. With respect to Eq.~\eqref{Kta}, here some additional
exponential are present which obstructs the reconstruction of the
initial wave function. So, if we put a second grating at distances
multiples of $2D^2/\lambda$, even for a perfect alignment between the
two gratings the wave function is partially stopped and thus the
intensity detected will be lower than in the free case. Similarly, for
a displacement between the gratings of an half period, a portion of
signal, even little, may be detectable further on. In such a scenario,
a decrease of the coherence length $\lco$, i.e., a growth of the
incoherence of the beam, leads to a progressive reduction of the
visibility of the total intensity on the screen, in agreement with the
behaviour of the experimental data reported in \cite{ztalbot2}. (An
improvement of this analysis should presumably take
into account: 1) the three free-standing gratings for the Talbot-Lau
interferometer effectively used in experiments; 2) a proper description
of the van der Waals interaction with the grating).

\end{widetext}

\subsection{\label{newp}Near-field interferometry and randomness of
  arrival times}

Near-field interferometry, such as Talbot-Lau interferometry, should
allow us to probe quantum effects also due to the motion along the
longitudinal direction, which so far has been treated as classical.
Such a treatment has been of course completely motivated by the
experimental conditions considered so far for which both \eqref{dpp}
and \eqref{sL} are satisfied with a high degree of approximation. But
suppose that position spread in the longitudinal direction $\Delta y$
is not completely negligible with respect to $L$. Then the arrival
times would have statistical fluctuations of order
$$
\Delta \tf \sim \Delta y/v.
$$
For the fullerene experiments in the Fraunhofer region such
fluctuations are not appreciable: in this case, $\Delta \tf \sim~\Delta
y/v\approx 5\times 10^{-9}\,\mathrm{s}$, and since $\tf\approx
6\times 10^{-3}\,$s, we have
\begin{equation}\label{ciccio}
\Delta \tf \ll \tf.
\end{equation}

Near-field interferometry, possibly with light particles, should be
able to test the measured intensity when \eqref{ciccio} is violated.
A first prediction is immediately suggested by \eqref{I1}: the
measured intensity is obtained by the intensity distribution consider
so far, namely $\rho(x,x,\tf)$, by convolution with $|\phi_0|^2$.
Thus, randomness of the arrival times appears as an independent noise
on the standard interference profile, which reduces the fringe
visibility as it were an additional source of ``decoherence''. This
effect could be confused with a sort of {\it intrinsic} decoherence
(in this regard see a recent proposal concerning atomic diffraction
by standing light wave \cite{bonifacio}).

More generally, one may analyze the predictions of \eqref{curint} in
the mesoscopic regime~\cite{nino}.  Let us consider a monochromatic
beam, devoid of angular divergence, composed by free light particles
of mass $m$. Let the beam be diffracted by two slits of width $d$ and
distance $D$.  It is convenient to describe the split wave function
$\Psi({\bf r})$ by means of two-bidimensional Gaussian wave packets,
whose barycenters move parallel to the optical axis $y$ with the
velocity $v=\hbar k_y/m$:
\begin{equation}\label{gaussian}
\begin{split}
  \Psi({\bf r})& = \psi_0(x) \phi_0(y)\\
  & = C\bigg[\sum_{s=\pm}e^{-\frac{(x-sD/2)^2}{4\sigma_x^2}}\bigg]\,
  e^{-\frac{y^2}{4\sigma_y^2}\,+\,ik_yy}.
\end{split}
\end{equation}
The constant $C$ ensures normalization. The transversal standard
deviation $\sigma_x$ of each wave packet is related to the slit width
$d$ (a typical assumption is $6\sigma_x=d$), while the longitudinal
one is related to the extension of the initial support of $\Psi({\bf
  r})$.

\begin{figure}[floatfix]
  \includegraphics [scale=0.6]{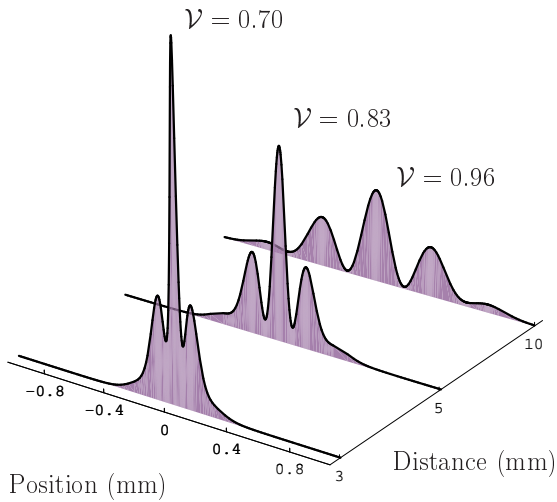}
\caption{
  Double-slit interference patterns for different values of the
  distance between the grating and the detection screen. Patterns are
  obtained for ultra-cold neutrons ($v\approx 1\,\mathrm{m/s}$)
  diffracted by a pair of slits of width $d=5 \times
  10^{-6}\,\mathrm{m}$ and separated by a distance
  $D=10^{-5}\,\mathrm{m}$. Neutron initial wave function is described
  by means of bidimensional Gaussian wave packets with a longitudinal
  standard deviation $\sigma_y= 10^{-3}\,\mathrm{m}$ and a transversal
  one $\sigma_x=d/6$.  }\label{fig3}
\end{figure}

After the splitting, we can consider that $\Psi({\bf r})$ evolves
according to the free Schr\"odinger equation \eqref{sch} until the
particle is detected on the detection screen placed at a distance
$L$. Given an arbitrary distance, not necessarily large compared with
the size of the support of $\Psi({\bf r})$, and for a free evolution,
the intensity detected at $y=L$ can be calculated by means of Eq.
\eqref{curnum}, with $\psi(x,t)$ and $\phi(x,t)$ obtained by the free
evolution of the initial state \eqref{gaussian}. The result of such a
numerical simulation for ultra-cold neutrons is shown in Fig.
\ref{fig3}. At distances as short as to be comparable with the
longitudinal spread of neutron wave packets, the statistical
fluctuations on arrival times produce an appreciable reduction of the
fringe visibility just as it would happen in case of an incoherent
preparation of the beam and/or in case of environmental-induced
decoherence. The only difference consisting in the distance
dependence of the different processes: this kind of ``decoherence''
reduces by increasing the distance, environmental-induced decoherence
increases, while the effects due to incoherent preparation are
independent.

In particular, notice that the reduced visibility is not due, not
even partially, to an incomplete wave-packet superposition, being
this ensured, also for the shortest distance shown in
Fig.~\ref{fig3}, by the large wavelength of ultra-cold neutrons.
Thus, the loss of fringe contrast has to be ascribed only to arrival
time fluctuations of the same order of the classical time of flight.

For simulations of Fig.~\ref{fig3} we used a longitudinal
delocalization at the double-slit given by $\sigma_y= 10^{-3}$\,m.
This assumption can be relaxed to shorter values still detecting
results analogous to those shown by Fig.~\ref{fig3}, but at shorter
distances. In this case the spatial resolution required for efficient
detection becomes higher.  Conversely, for a larger longitudinal
delocalization, intrinsic decoherence effects can be readily detected
at larger distances by means of less refined detectors.

In conclusion, let us underline that, since very slow neutrons
($v\approx 1$\, m/s) \cite{scheckenhofer} characterized by a wide
transversal support ($\Delta x \gtrsim 10^{-4}$\,m)
\cite{scheckenhofer, gahler} have been used in interferometry, we
think that an experimental test of the predicted behavior shown in
Fig.~\ref{fig3} might be indeed in the reach of present technology
(not necessarily for the concrete situation we have simulated, which
was mainly for illustrative purposes).

\section{\label{s5}Numerical calculations}

In this section we shall use Eq.~\eqref{N} to fit the experimental
data reported in~\cite{JMO2000}. In this regard, note that
Eq.~\eqref{N} describes an ideal situation where an infinitely
accurate detector measures the spatial intensity distribution of a
strictly monochromatic beam. Some adjustments have to be carried out
in order to include in our treatment the effects on the diffraction
pattern due both to the velocity distribution characterizing the beam
macromolecules and to the distortions unavoidably introduced during
the measurement process \cite{zeil88}. Each of these corrections have
to be implemented on the intensity level, since they consist in
incoherent contributions.

The total intensity is obtained by the sum of the monochromatic
components of the beam
\begin{equation}\label{57}
\bar{I}(x) = \int \ud \lambda\; f(\lambda)\,
I(x,\lambda),
\end{equation}
where $I(x,\lambda)$ is given by Eq.~\eqref{N} and its dependence on
$\lambda$ is shown by Eq.~\eqref{tf}. The wavelength distribution
$f(\lambda) \ud \lambda$ is directly obtained by the \em supersonic
velocity distribution \em $f(v) \ud v$ characterizing the
macromolecule ensemble~\cite{JMO2000}
\begin{equation}\label{supersonic}
f(v)\,\mathrm{d}v\,\propto\,v^3 \exp\,
\big[-(v-v_0)^2/\hat{v}^2\big]\,{\rm d}v\,.
\end{equation}
It describes beams in transition between effusive and slow jet
sources~\cite{scoles}.  The parameters $v_0$ and $\hat{v}$ depend both
on the temperature of the beam and on the physical features of the
given molecule and they are deduced by a best fit over experimental
measurement of the velocity distribution (see Fig.~2
in~\cite{JMO2000}).

\begin{figure}[h!]
  \includegraphics [scale=0.6]{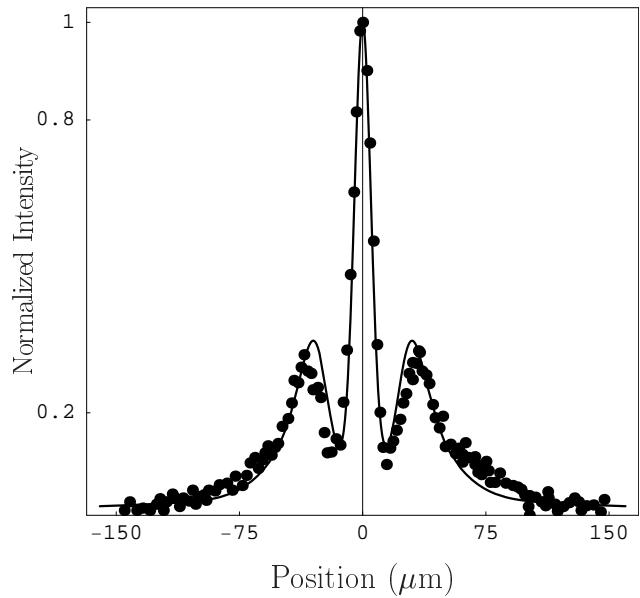}
\caption{Theoretical fit and experimental data for $\mathrm{C}_{60}$
interferometry
  \cite{JMO2000}. Theoretical fit was performed considering $N=10$
  equal, rectangular shaped, slits of an effective width
  $d=36\,\mathrm{nm}$.  The spatial resolution of the detector is
  $2\widetilde{x} =8\,\mu{\rm m}$. The theoretical curve and experimental
  data are normalized to the value of the central
  maximum.}\label{fig4}
\end{figure}

The finite spatial resolution can be taken into account by an
integration over the size of the elementary detector, say $2\widetilde{x}$,
weighed on its spatial response function $D(x)$.  In particular, in
the case of a flat response function, $D(x)$ becomes the
\emph{characteristic function} defined on the interval
$\big[-\widetilde{x}\,,\,+\widetilde{x}\big]$ and the {\it effective detected
  intensity} is expressed by a moving average
$$
I_{\mathrm{eff}}(x)\,=\,
\int_{-\widetilde{x}}^{+\widetilde{x}}\hspace*{-12pt}\ud \zeta\:\bar{I}(x+\zeta)\:D(\zeta)
\,=\, \frac{1}{2\widetilde{x}}\int_{x-\widetilde{x}}^{x+\widetilde{x}}\hspace*{-17pt}\ud
\zeta\ \bar{I}(\zeta),
$$
whence, from Eq.~\eqref{57},
\begin{equation}\nonumber
I_{\mathrm{eff}}(x)\
\,\propto\,
\int_{x-\widetilde{x}}^{x+\widetilde{x}}\hspace*{-15pt}\ud \zeta \int\! \ud \lambda\;
f(\lambda)\,I(\zeta,\lambda).
\end{equation}

\begin{figure}[h!]
  \includegraphics [scale=0.6]{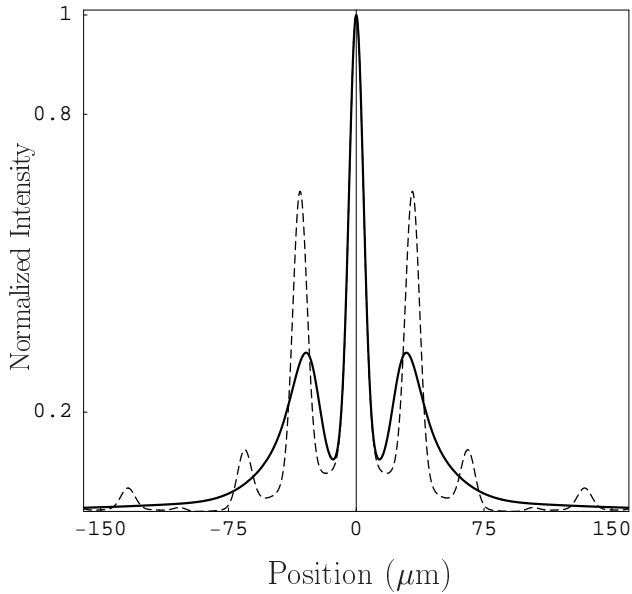}
\caption{Interference patterns due both to a macromolecular beam
  characterized by a velocity distribution given by
  Eq.~\eqref{supersonic} as in~\cite{JMO2000} (full line) and to a
  strictly monochromatic beam corresponding to the mean velocity
  (dashed line).  The curves are normalized to the value of the
  central maximum.}\label{fig5}
\end{figure}

\begin{figure}[floatfix]
  \includegraphics [scale=0.5]{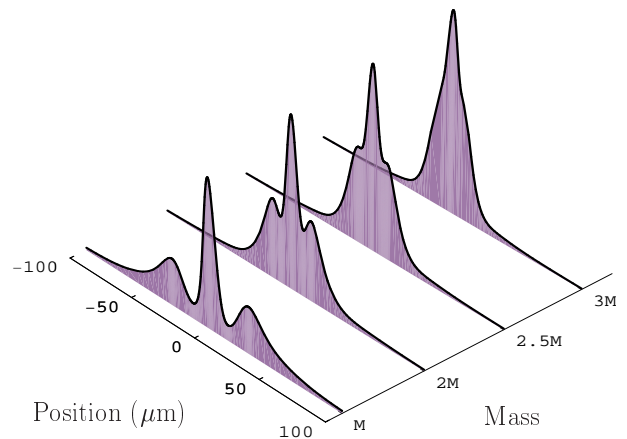}
\caption{Diffraction of molecules with masses multiple of the $C_{60}$
  mass $M$. For larger masses the quantum behavior becomes
  progressively negligible, approaching to the classical limit. Every
  curve has been normalized to the value of the central maximum
  obtained for the mass $M$.}\label{fig6}
\end{figure}

\begin{figure}[floatfix]
  \includegraphics [scale=0.5]{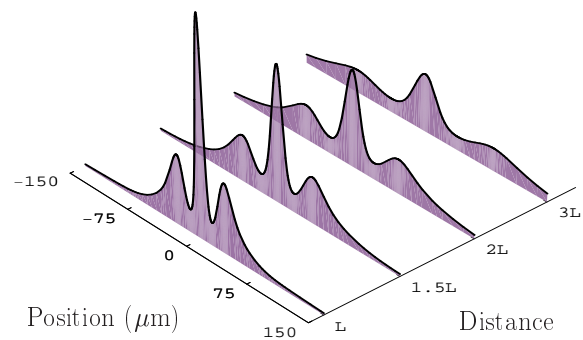}
\caption{Interference patterns for different values of the distance between
  the grating and the detection screen, in unit of the value $L$
  reported in~\cite{JMO2000}. Increasing the distance the pattern
  spreads in position with respect to the optical axis. The visibility
  is reduced by the increased number of decohering events.  Every
  curve has been normalized to the value of the central maximum
  obtained for the distance $L$.}\label{fig7}
\end{figure}

\begin{figure*}[floatfix]
  \includegraphics* [scale=1.1]{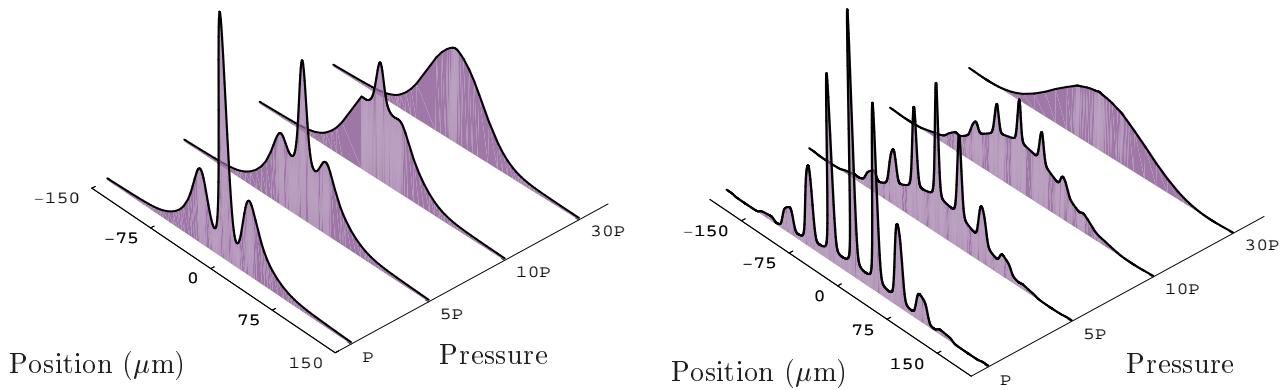}
\caption{The two figures show interference patterns for increasing
  pressures at fixed temperature, in unit of the pressure $P=5\times
  10^{-6}$Pa used in~\cite{JMO2000}. Left figure corresponds to the
  experimental setup reported in~\cite{JMO2000}: slit effective width
  $d=3.6\times 10^{-8}$\,m and collimator aperture $a=10^{-5}$\,m.
  Right figure has been obtained for narrower slits ($d=2\times
  10^{-8}$\,m), an improved collimation ($a=5\times 10^{-6}$\,m) and a
  velocity selected $C_{60}$ beam ($\Delta v/v = 10\%$, where $v$ is
  the mean velocity).  Every curve has been normalized to the value of
  the central maximum at the pressure $P$.}\label{fig8}
\end{figure*}

For the free gap $d$ of the rectangular slits, it was used, according
to~\cite{grisenti}, the effective width estimated in
Refs.~\cite{arnzei99, JMO2000} as previously discussed in
Section~\ref{incoherence} \footnote{In other words, we have taken the
  $\varphi _s(x)$'s in \eqref{inco} to be characteristic functions
  with supports on the effective width of the slits. Numerical
  simulations show stability with respect to other equivalent choices
  for these functions.}. Moreover, a constant background has been
subtracted from experimental data~\cite{arnzei99,JMO2000}.

Finally, referring to Eq.~\eqref{N}, note that the number $N$ is
basically {\em a priori} unknown, being the effective number of slits
which concretely contribute to diffraction. In fact, only the
knowledge of the fullerene wave function on the grating would allow
us to fix the effective number of slits in which the initial wave
function is split, since the width of collimation only gives
information about the maximum number of wave packets which could
contribute to the interference pattern. Nevertheless, $N$ can be
easily inferred {\em a
  posteriori} as a free parameters of the fit with experimental data.
So doing, we find $N=10$.

Our theoretical evaluation for the interference pattern is shown in
Fig.~\ref{fig4}, together with the experimental points published in
Ref.~\cite{JMO2000}, page 2819, Fig.~5.

\section{Conclusions and perspectives}\label{simulations}

One of the main goals of our work has been that of deducing, within
the framework of Joos and Zeh model, a geometric optics limit of
quantum mechanics in the presence of decoherence.

Our theoretical analysis confirms the negligibility of environmental
disturbances in recent experiments of macromolecular interferometry
\cite{arnzei99, JMO2000} with respect to the loss of coherence due to
beam production.

Moreover, our analysis reproduces classical results, as well as
Fraunhofer relation for optical diffractive patterns, and provides
the quantum-mechanical analog for interferometry with partially
coherent sources of radiation. For the latter subject, referring to
the theory of partial coherence, existing analogies can be pointed
out in a deeper way by exploiting numerical simulations.
Fig.~\ref{fig5} clearly shows that, by selecting the particle beam
velocity, the visibility of the corresponding interference pattern
does not tend to unity, being upper bounded by the damping term
$\mathcal{V}_\mathrm{QM}$ (cf. Eqs.  \eqref{dt}--\eqref{vis}). On the
other hand, the effect of the velocity selection makes more
interference fringes visible at the border of the interference
pattern.

The same behavior is obtained in the framework of classical optics,
studying interference patterns due to quasimonochromatic
fields~\cite{wolf1, wolf2}. Classical partial coherence theory,
supported by recent experiments~\cite{otto2}, states that, by
filtering, the pattern visibility at most approaches the value of the
modulus of the \emph{ degree of spectral coherence} $\mu (\lambda)$,
which depends on the \emph{spatial coherence} of the source, and that
more fringes becomes visible, since the {\em temporal coherence} is
improved.

Moreover, our approach provides a useful theoretical framework for
analyzing present (and possibly new) interference experiments. For
instance, we have studied the mass dependence of the interference
pattern due to the diffraction of heavy particles. Fig.~\ref{fig6}
shows the simulations corresponding to beams of macromolecules
heavier than $\mathrm{C}_{60}$, but characterized by the same
physical features. Note that for larger masses the quantum behavior
becomes progressively negligible, approaching to the classical limit.
There are, indeed, other experimental investigations which support
the previous expected behavior\cite{arndt01}. They show that
visibility of $\mathrm{C}_{70}$ diffracted beams is slightly reduced
than that obtained for $\mathrm{C}_{60}$ ones in the same conditions.

In studying the effects of decoherence, it is particularly
interesting to analyze the case of a diffracted quantum particle
which experiences a lot of scattering events before reaching the
detection screen. This situation is realized for larger
grating-screen distances (see Fig.~\ref{fig7}), or for more frequent
scattering processes due, for example, to increasing value of
pressure (see Fig.~\ref{fig8}).  In both cases the fringe visibility,
and thus the wavelike behavior of the molecule, is progressively
corrupted.

By improving the collimation of velocity selected $\mathrm{C}_{60}$
beams, the interference pattern shows a richer structure of fringes
and thus a more evident quantum behavior.  Moreover, interference
oscillations appear also in less restrictive environmental
conditions, provided that the signal-to-noise ratio is such to allow
experimental detection. In fact, as shown by the right plot in
Fig.~\ref{fig8}, side maxima, far from the optical axis, which are
not detected at pressure $P$ in experiments~\cite{arnzei99,JMO2000},
turn out to be clearly visible even for pressures ten times larger
than $P$. This might be relevant in devising new interferometry
experiments directed to the study of the quantum behavior of
macroscopic objects and also to test quantitatively the effects on a
quantum subsystem due to external noise. Our suggestion has been
partially realized in very recent experiments~\cite{nairz03}, which
show results in agreement with the prediction of Fig.~\ref{fig8}.

\begin{acknowledgments}
  We thank G.~Dillon for an initial stimulating observation, L.~Basano
  and P.~Ottonello for continuous and fruitful discussions and
  D.~D\"urr for useful comments. Finally, we thank B.~Vacchini for a
  careful reading of an earlier draft of this manuscript and for
  helpful suggestions. This work was financially supported in part by
  INFN.
\end{acknowledgments}

\appendix

\begin{widetext}
\section{Young interference pattern with Gaussian slits}\label{Y}

In the following we shall develop the exact solution of~\eqref{ro} in
the case of slits with a Gaussian shaped profile. For simplicity we
shall treat the case of interference patterns due to just a pair of
slits of width $d$ and distance $D$, even though an analytical
solution can be obtained also for a grating composed of several
slits.

Let us consider an initial wave function split by two Gaussian slits
of standard deviation $\sigma_{x}$ and centered at $x=\pm D/2$ (we
can typically choose $6\sigma_x =d$)
\begin{equation}\label{psiapp}
\psi_0 (x;k_x)=\bigg[
\varphi\Big(x+\frac{D}{2}\Big)+\varphi\Big(x-\frac{D}{2}\Big)\bigg]\,
e^{i\,k_x\,x}
=C\:\bigg [e^{-\frac{(x-D/2)^2}{4\sigma_{x}^2}} +
e^{-\frac{(x+D/2)^2}{4\sigma_{x}^2}}\bigg]\,e^{i\,k_x\,x}\,,
\end{equation}
where $C$ is the normalization constant.  Inserting
Eq.~\eqref{rotilde} in Eq.~\eqref{ro} with the initial
state~\eqref{psiapp} we get
\begin{equation}\nonumber
\begin{split}
  I(x)=\frac{MC^2}{2\pi\hbar\tf }
  \iint^{+\infty}_{-\infty}\!\!\!\!\!\!\!\!\!{\rm d} x_0\,{\rm d}
  x_0'\, &\exp\bigg\{\frac{iM}{2\hbar \tf
    }\:\big[x_0^2-x_0'^2+2x(x_0'-x_0)\big]-\frac{(x_0-x_0')^2}{2\lcoq}
  \bigg\}\\
  \times\sum_{s,s'=\pm} &\exp\bigg
  [-\frac{\big(x_0+s\frac{D}{2}\big)^2}{4\sigma_{x}^2}-\frac{\big(x_0'+s
    '\frac{D}{2}\big)^2}{4\sigma_{x}^2}\bigg ]\, ,
\end{split}
\end{equation}
whence
\begin{equation}\label{2G}
I(x)\:=\:
\frac{4M\sigma_{x}^2C^2}{\hbar \tf \sqrt{Q(\tf )}}\,
\exp\bigg[-\frac{x^2+D^2/4}{Q(\tf )}\bigg(\frac{\sqrt{2}M \sigma_{x}}
{\hbar \tf }\bigg)^{\!2}\,
\bigg]
\bigg\{\cosh\bigg[\bigg (\frac{\sqrt{2}M\sigma_{x}}{\hbar
\tf }\bigg )^2\frac{Dx}{Q(\tf )}\bigg]+\exp\bigg[-\frac{D^2}{2Q(\tf )\lcoq}
\bigg]\cos \bigg
[\frac{M D x}{\hbar \tf  Q(\tf )}\bigg]\bigg\},
\end{equation}
where $Q(\tf )\equiv 1+\big[2M\sigma_{x}^2/(\hbar\tf)\big]^2 +
(2\sigma_{x}/\lco)^2$.

For $\tf \gg M\sigma_{x}D/\hbar$ and $\lco\gg \sigma_{x}$ the term
$Q(\tf )\to 1$, the first exponential in Eq.~\eqref{2G} reduces to
$\exp\big\{\!\!-~\!\!\![\sqrt{2}Mx\sigma_{x}/(\hbar\tf )]^2\big\}$, whence
$x_{max}\approx \hbar \tf / (M \sigma_{x})$, and thus the argument of
the hyperbolic cosine is close to zero. It follows that Eq.~\eqref{2G}
is well approximated by (see Eq.~\eqref{tf})
\[
I(x)\:=\: \frac{8\pi\sigma_{x}^2C^2}{\lambda L}\:
\exp\bigg[\bigg(\frac{2\sqrt{2}\pi x\sigma_{x}}{\lambda L}\bigg)^{\!2}\,
\bigg]\bigg [1+e^{-\frac{D^2}{2\lcoq}}\:\cos \bigg(\frac{2\pi
  Dx}{\lambda L}\bigg) \bigg ]\:=\:\frac{4\pi}{\lambda L}\,
\bigg\vert\widehat{\varphi}\bigg( \frac{2\pi x}{\lambda L} \bigg)
\bigg\vert ^2 \,\bigg [ 1+e^{-\frac{D^2}{2\lcoq}}\:\cos \bigg (
\frac{2\pi D x}{\lambda L} \bigg )\bigg ]\,,
\]
which coincides with Eq.~\eqref{N} evaluated for $N=2$.\\

\end{widetext}

\section{Approximation quality test for fullerene
  experiment}\label{aQ}

In this appendix we evaluate the error committed introducing the
zero-order approximation of \eqref{bigapp} in Eq.~\eqref{change},
referring to experimental conditions reported in~\cite{JMO2000}.

Since $\vert (s'-s)D\vert\lesssim 2 \lco$, we will test the previous
approximation in the most unfavorable situation, according to
$\exp[(\xi'-\xi)/\lco]\approx 1$. To this aim, it is useful to
introduce the integrals
\begin{eqnarray}\nonumber
\mathcal{I}(\lco,\bar{k})=\int_{-\infty}^{+\infty}\!\!\!\!\!{\rm d}\xi\
\varphi(\xi)\,e^{\xi/\lco}\, e^{i\bar{k}\xi}\,, \\\nonumber
A(\bar{k})=\int_{-\infty}^{+\infty}\!\!\!\!\!{\rm d}\xi
\:\varphi(\xi)\,e^{i\bar{k}\xi}\,,
\end{eqnarray}
where $\bar{k}=M x/(\hbar \tf)$. Zero-order approximation of
\eqref{bigapp} can be checked by evaluating the relative displacement
of the square modulus of the previous integrals
$$
R(\lco,\bar{k})=\bigg\vert\frac{\vert
  \mathcal{I}(\lco,\bar{k})\vert ^2- \vert A(\bar{k})\vert ^2}{\vert
  \mathcal{I}(\lco,\bar{k})\vert ^2}\bigg \vert\,.
$$
The less is the value assumed by $R(\lco,\bar{k})$, the better is
the quality of the zero-order approximation of \eqref{bigapp}.

By making explicit $\varphi (\xi)$ with the characteristic function
defined in the interval $[-d/2, d/2]$, a straightforward calculation
leads to
\begin{equation}\nonumber
\begin{split}
  \vert \mathcal{I}(\lco,\bar{k}) \vert ^2 &\ =\ \frac{2}{1/\lco^2
    +\bar{k}^2}\,[\cosh(d/\lco)-\cos(\bar{k}d)],\\
  \vert A(\bar{k})\vert ^2 &\ =\ d^2\,{\rm sinc}^2 (\bar{k}d/2)\ =\
  \frac{2}{\bar{k}^2}\,[1-\cos (\bar{k}d)],
\end{split}
\end{equation}
whence
\begin{equation}\label{R}
R(\lco,\bar{k})=\bigg\vert 1-\bigg [1+\frac{1}{(\lco\bar{k})^2}\bigg ]
\frac{1-\cos(\bar{k}d)}{\cosh(d/\lco)-\cos(\bar{k}d)}\bigg\vert\,.
\end{equation}
In the limit of small $\bar{k}$ (i.e., for positions $x\approx 0$,
close to the maximum of intensity), we get
\begin{equation}\label{1p}
\lim_{\bar{k}\to 0}R(\lco,\bar{k})=1-\frac{(d/\lco)^2}{2\,
[\cosh(d/\lco)-1]}\sim\frac{(d/\lco)^2}{12} \approx 0.011\,.
\end{equation}
This means that in correspondence of the greatest intensities the
approximate expression just moves away from the correct one of about
1\%.

Since $R(\lco,\bar{k})$ is an increasing function of $\bar{k}$, and
thus of $\vert x\vert$, then the most unfavorable case takes place at
the edge of the interference pattern, where, however, the intensity is
negligible.  By using $\bar{k}_{\rm max}\approx d^{-1}$ in evaluating
\eqref{R}, it results that $R(\lco,\bar{k}_{\rm max})$ does not
significantly differ from evaluation \eqref{1p}.

\end{document}